\documentstyle[12pt,aasms4]{article}

\lefthead{F. J. Masci et al.}
\righthead{A New Complete Sample of Sub-millijansky Radio Sources}

\begin{document}

\title{A New Complete Sample of Sub-millijansky Radio Sources: An Optical 
and Near-Infrared Study}

\author{Frank J. Masci\altaffilmark{1,2},
J. J. Condon\altaffilmark{3}, 
T. A. Barlow\altaffilmark{1}, 
C. J. Lonsdale\altaffilmark{1},
C. Xu\altaffilmark{1},
D. L. Shupe\altaffilmark{1}, 
O. Pevunova\altaffilmark{1,4}, 
F. Fang\altaffilmark{1} and 
R. Cutri\altaffilmark{1}}

\altaffiltext{1}{Infrared Processing and Analysis Center, M/S 100-22,
California Institute of Technology, Jet Propulsion Laboratory, 
Pasadena, CA 91125}
\altaffiltext{2}{Electronic mail: fmasci@ipac.caltech.edu.au}
\altaffiltext{3}{National Radio Astronomy Observatory, 520 Edgemont Road, 
Charlottesville, VA 22903} 
\altaffiltext{4}{Present address: Interferometry Science Group, M/S 171-122C,
Jet Propulsion Laboratory, Pasadena, CA 91109-8099} 

\begin{abstract}
The Very Large Array (VLA) has been used in C-configuration to map an
area $\simeq0.3\,{\rm deg}^{2}$ at 1.4 GHz with $5\sigma$
sensitivities of 0.305, 0.325, 0.380 and
0.450 mJy beam$^{-1}$ over four equal subareas. 
Radio properties are presented for 62 detected
sources.  Deep optical imaging to Gunn $r\simeq25$ mag using the Hale
5-m telescope covering $\simeq0.21\,{\rm deg}^{2}$ is reported for a
subset of 43 sources.  
This optical follow-up is much deeper than that of existing larger area
radio surveys of similar radio sensitivity.
Archival $J$, $H$ and $K$-band photometry from
the Two-Micron All Sky Survey (2MASS) is also presented.  Using a
robust likelihood-ratio technique, we optically identified 26 radio
sources with probabilities $\gtrsim80\%$, nine are 
uncertain/ambiguous detections, and eight are empty fields.
Comparisons with a stellar synthesis model that includes radio
emission and dust reddening suggests that the
near-infrared--to--optical emission in a small, bright sub-sample is
reddened by `optically thin' dust with absorption
 $A_{V}\simeq2-2.5$ mag,
regardless of morphological type.  
This is consistent with other, more
direct determinations of absorption.
The radio--to--optical(--near-infrared) flux
ratios of early-type galaxies require significant contamination in
the radio by an active galactic nucleus (AGN), consistent with the
current paradigm. Using our simple
modeling approach, we also discuss a potential diagnostic for
selecting Ultraluminous Infrared Galaxies (ULIGS) to $z\simeq1.6$ from
microJansky radio surveys.
\end{abstract}


\section{INTRODUCTION}

The advent of deep radio surveys reaching flux densities well below 1
mJy (Mitchell \& Condon 1985; Windhorst et al. 1985; Oort 1987;
Windhorst et al. 1993; Hopkins et al. 1998; Richards 2000) revealed a
new population of faint sources more numerous than the AGN-powered
radio galaxies dominating the strong-source population. This
corresponds to a steepening of the differential source counts 
over non-evolving predictions at levels
$\lesssim4$ mJy. 
The faint counts suggest that significant evolution has
occured over the redshift range spanned by the observed population.
Photometric and spectroscopic studies (Thuan et al. 1984; Windhorst et
al. 1985; Thuan \& Condon 1987; Benn et al. 1993) suggest that the
faint excess at 1.4 GHz is composed predominately of star-forming
galaxies similar to the nearby starburst population dominating the
{\it Infrared Astronomical Satellite} ({\it IRAS}) 60-$\mu$m counts
(Benn et al. 1993).  Indeed, this is supported by the strong
correlation between radio (1.4 GHz) and far-infrared (60-$\mu$m) flux
densities of disk galaxies (Helou, Soifer \& Rowan-Robinson 1985), implying
a significant proportion of starburst galaxies at faint radio flux
densities.
 
The overall observed source-count distribution from faint ($\mu$Jy) to 
bright ($S_{1.4}\gtrsim10$mJy) flux densities cannot be explained by 
starbursts alone. Evolutionary models of radio source counts need to
invoke two separate populations (eg. Danese et al. 1987; 
Rowan-Robinson et al. 1993; 
Hopkins et al. 1998). 
Condon (1989) describes these
populations as starbursts and monsters, each powered by different 
mechanisms: `starbursts' deriving their radio emission from supernova 
remnants and H{\small II} regions and `monsters' from compact 
nuclear related activity (eg. active galactic nuclei; AGNs). 
The proportion of AGN is much greater at higher flux densities  
$S_{1.4}\gtrsim10$ mJy
(Kron, Koo \& Windhorst 1985; Gruppioni et al. 1998), where
a majority are associated with classical radio galaxies exhibiting 
extended (FRI and FRII-type) morphologies (Fanaroff \& Riley 1974). 
The optical counterparts to sources at bright
radio flux densities $S_{1.4}\gtrsim1$ mJy is
comprised mostly of ellipticals
while at sub-mJy to $\mu$Jy levels, the optical counterparts are identified 
as blue galaxies exhibiting peculiar (compact, 
interacting and irregular) morphologies (Kron, Koo \& Windhorst 1985;
Gruppioni et al. 1998).
Studies of faint radio sources, 
namely their stellar population, how they 
evolve with redshift, and how they relate to local normal galaxies is
progressing rapidly, however, much remains to be learned from the
faintest ($\mu$Jy) radio populations at redshifts $1\lesssim z\lesssim2$. 

Radio surveys are insensitive to the effects of absorption 
by dust which is known to bias surveys severely
at optical/UV wavelengths. This is particularly important for derivations
of the cosmic history of star formation and its relation to hierarchical
models of galaxy formation. Optical/UV studies have shown that there is
an increase in both the space density of star-forming, morphologically
disturbed galaxies (eg. Richards et al. 1999)
and also, the global star formation rate with 
redshift to $z\gtrsim1$ (eg. Madau et al. 1996). Similar evidence
is emerging from studies of sub-mm sources (Blain et al. 1999)
and amongst the faint radio
population at $1\lesssim z\lesssim2$ - 
the redshift range probed by the deepest surveys
(eg. Cram 1998; Haarsma et al. 1999). 

There has been much speculation as to whether global star-formation 
rates (SFRs) derived from radio observations exceed those 
determined from optical/UV studies.
Cram et al. (1998) note that systematic
discrepancies may exist between the various star formation indicators,
which are not well understood (see also Schaerer 1999).
It is encouraging to see however that Haarsma et al. (2000)
find global SFRs derived from 1.4GHz observations of the Hubble Deep Field
to exceed optically detetermined values by a factor of a few out to  $z\sim1$.
Indeed, an analysis
of Balmer decrements and optical--near-infrared colors in star forming 
galaxies by 
Georgakakis et al. (1999) 
from the Phoenix Deep Survey 
(Hopkins et al. 1998), finds evidence for visual extinctions from one to a few
magnitudes. Currently, about 20\% of existing micro-jansky radio samples
remain unidentified to $I = 25$ mag in Hubble Deep Field images
(eg. Richards et al. 1999). A majority of these could
represent
a significant population of dust-enshrouded starbursts and/or AGN at high
redshift. 
These results are in support of efforts to further understand the
dust properties of star-forming galaxies.
 
The primary aim of this paper is to introduce
a new complete sample of radio sources selected at 1.4 GHz, 
uniformly selected over the flux range $S\gtrsim0.3$ mJy (5$\sigma$) from an 
area covering $\simeq 0.3\,{\rm deg}^{2}$. Although much larger area surveys
to deeper radio sensitivities have been carried out 
(eg. Hopkins et al. 1998), the present study reports the results of
more sensitive optical observations.
Archival near-infrared data for a subset of the sample are also presented. 
The near-infrared data is from the ongoing 2MASS project, and represents
a unique aspect of this study in the identification 
of radio-selected starbursts. 
Although we currently lack valuable spectroscopic information, 
we combine radio--near-infrared--optical flux ratios, radio maps, 
and optical images to explore
the properties of the entire sample.
Our deep optical identifications provide the 
opportunity to asses the importance of dust in star-forming systems
via the observed radio--to--optical and near-infrared--to--optical colors.
Simple stellar synthesis models that include radio emission and reddening 
are used to constrain possible amounts of absorption.

This paper is organised as follows. In \S 2, we discuss the radio 
observations and data reduction, present the radio catalog, and 
compare our results with data available from (shallower) all sky radio surveys.
The optical photometric observations, their reduction
and astrometric calibrations are discussed in \S 3. 
Our method for radio-optical
identification, the archival near-infrared data, radio-optical image
overlays, and our optical/near-infrared catalog are presented in \S 4. 
A study of the radio, near-infrared, and optical colors
and constraints
on synthesis models incorporating dust is presented in \S 5.
An application of our color-color analysis to select high-redshift
ultraluminous infrared galaxies from deep radio surveys is discussed in \S 6. 
All results and future prospects are summarised in \S 7.

\section{RADIO OBSERVATIONS}

\subsection{Strategy}
\label{strat}

Observations were made with the VLA C-configuration at 
1.4 GHz on 1998 December 19. This configuration yields a good
compromise
for resolution and surface-brigntness sensitivity.
The $5\sigma$ confusion limit for
this configuration is only $50\mu$Jy/beam at 1.4 GHz 
(Mitchell \& Condon 1985) since the synthesised
beam size (Full Width at Half Power; FWHP) is $\sim15$ arcsec.  
The resulting radio positions have rms errors $\sim1$ arcsec,
except for extended 
sources with multiple components, sufficient for making optical
identifications. 
At 1.4 GHz the FWHP of the
VLA primary beam is 31 arcmin and approximately corresponds to the diameter of
our final imaged field. 
This relatively large coverage avoids
field-to-field variations in source counts induced by high-redshift 
clustering.
Although surveys at higher frequencies (eg. 8 GHz) 
can reach lower flux densities 
than at 1.4 GHz, most radio sources have spectral indices $\alpha\sim1$
($S\propto\nu^{-\alpha}$) so that the population being sampled is
similar.

To provide uniform sensitivity over the full area of our field, we observed 
seven positions arranged in a filled hexagonal pattern 
with a separation of $26\arcmin$ between pointing centers.
The resulting root mean square (rms) map noise
is thus nearly constant (cf. Condon et al. 1998).
Our  field was centered on 
RA(2000)=$00^h 13^m 12^s.0$, Dec.(2000)=$+25\arcdeg 54\arcmin
44\arcsec$. 
This field was chosen for its relatively low foreground galactic-cirrus 
emission likely to affect optical/near-infrared identifications and
also for the absence of bright radio galaxies.
The integration time on each pointing was $\sim1$ hour. This allowed us to
reach an rms noise of $\sim60\mu$Jy in regions free from bright
contaminating sources (see \S~\ref{radnoise} for more details).

Our observations were made in spectral line mode 
with four Intermediate Frequencies (IFs), each divided into
7 spectral channels of width 3.125 MHz. The advantage of this mode is 
to minimize
bandwidth smearing (i.e. chromatic aberration) which reduces 
the point-source sensitivity away from the pointing center and
causes appreciable image distortion. 
Additionally, the spectral line mode is less prone to narrowband
interference noise spikes.
With continuum mode however, we would have had a little over
twice the bandwidth and a factor $\approx\sqrt{2}$ lower noise.

\subsection{Data Reduction}
\label{radred}

The data were analysed with the NRAO {\tt AIPS} reduction package. 
We observed 3C 48 to calibrate the visibility amplitudes, using $S =
16.5$ and $S = 15.9$ Jy at 1.365 and 1.435 GHz, respectively.  
The calibration was applied using the {\tt SPLIT} task.
The
calibrated data from each pointing were edited and imaged separately,
{\tt CLEAN}ed, and restored with 15 arcsec FWHP Gaussian beams.  The seven
separate images were weighted and combined as described in Condon et
al. (1998) to produce a final $33\arcmin\times33\arcmin$ map 
with nearly uniform
sensitivity and corrected for primary-beam attenuation.

\subsection{Noise and Source Extractions}
\label{radnoise}

\begin{figure}
\plotone{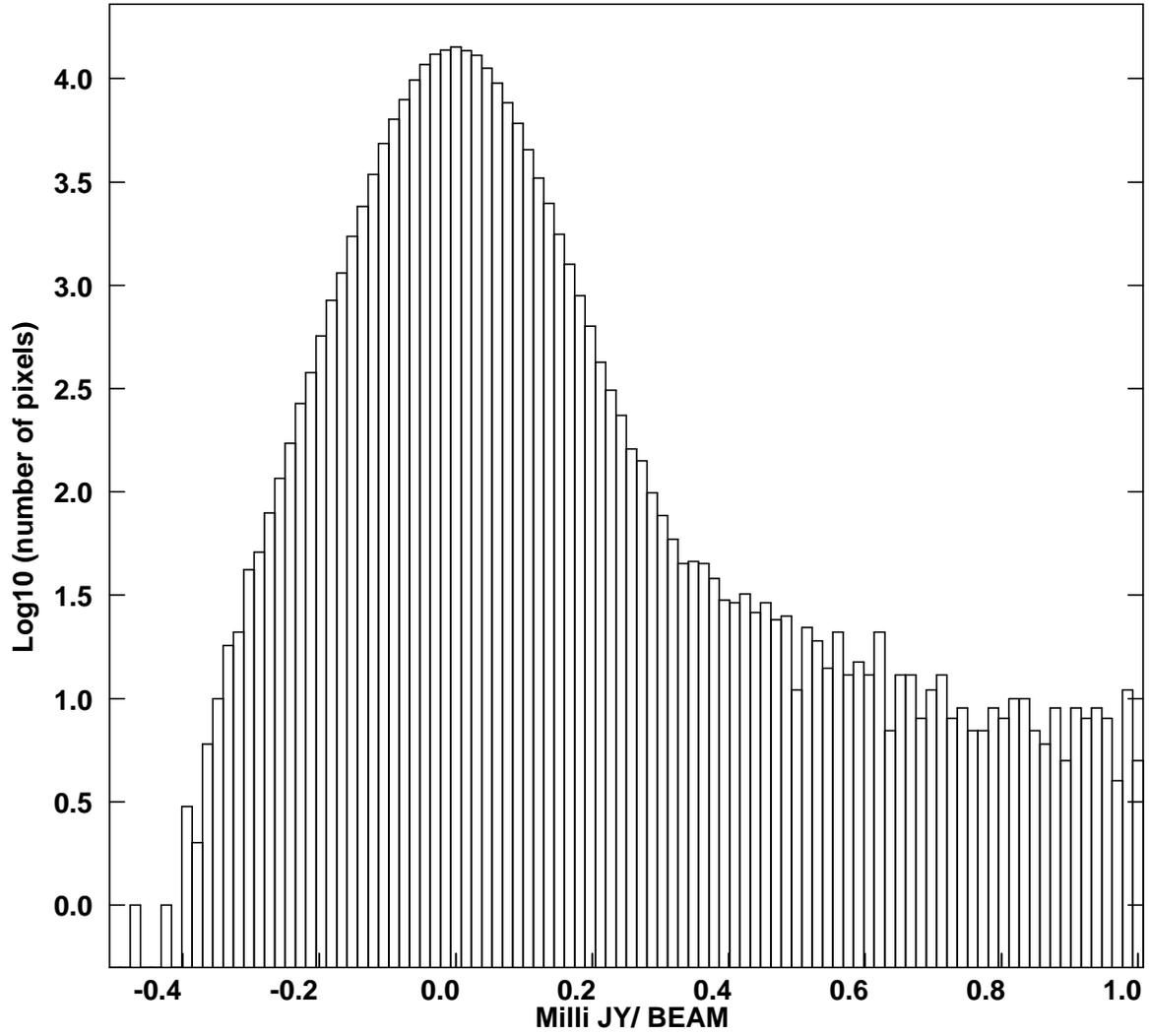}
\caption{Logarithmic histogram of pixel flux densities in our
complete radio map.
\label{pixhist}}
\end{figure}

\begin{figure}
\plotone{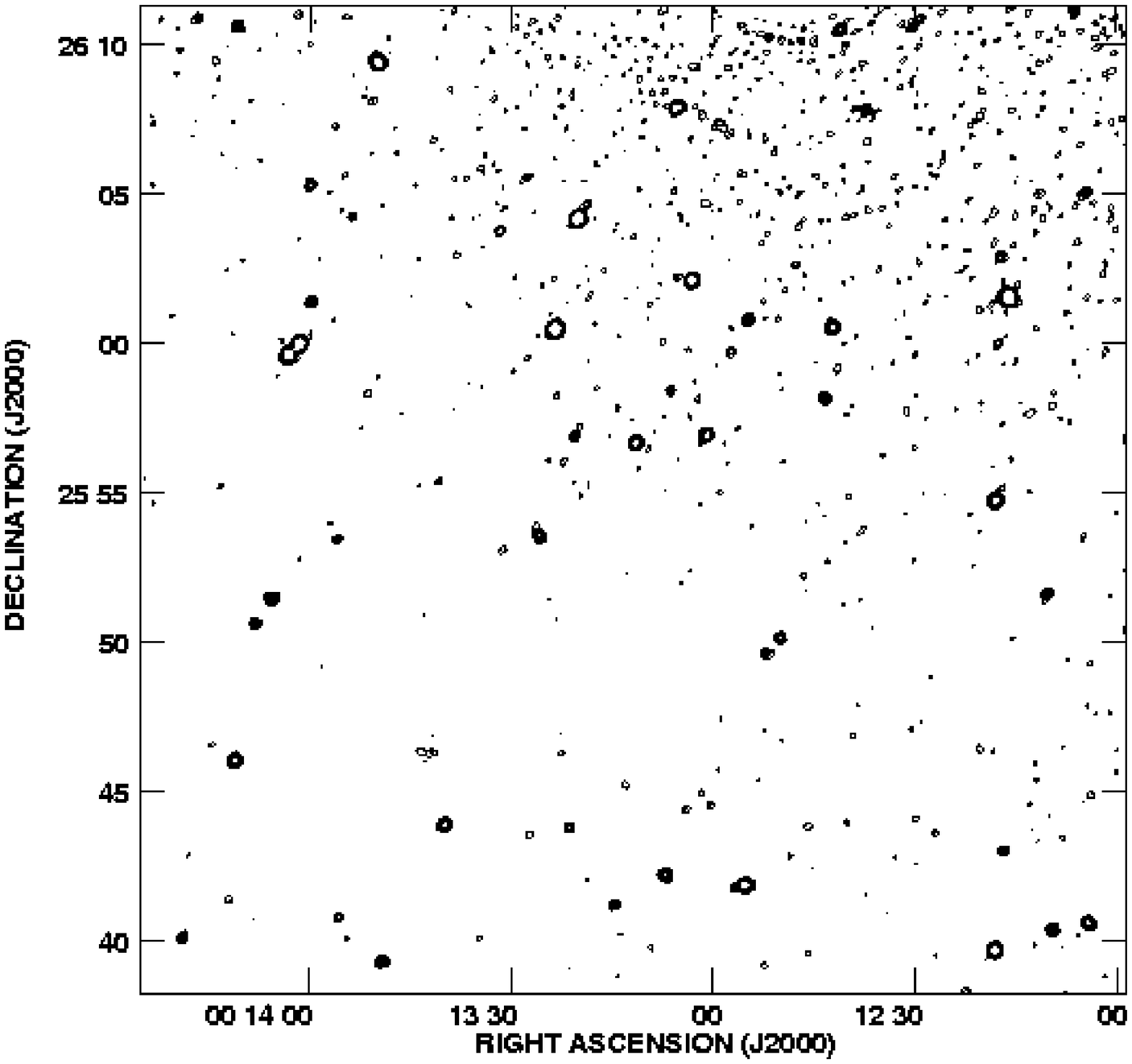}
\caption{Contour plot of our radio map covering
an area $33\times33\,{\rm arcmin}^{2}$ = $495\times495$ pixels.
Contour levels are 0.2, 0.4, 0.6, 0.8, 1.0, 1.2, 1.4, 1.6 mJy/beam.
Note the increase in rms noise from
the bright off-field source in the top right corner.
\label{contfull}}
\end{figure}

The resulting rms noise of our final map after correcting for
primary-beam attenuation is not uniform over the entire field,
but increases by up to 30\% near a strong ($\gtrsim400$ mJy) source
near the edge of our field.  Despite this variation in sensitivity, we
were able to divide our $33\times33\,{\rm arcmin}^{2}$ field
into four equal ($\approx16.5\times16.5\,{\rm arcmin}^{2}$) regions
within which the rms noise varies by no more than a few percent.
These constant-noise regions simplify the application of an automated
source-finding algorithm over a single continuous region (see below).
The lowest and highest rms noise amongst these regions was
$\simeq60.9\,\mu$Jy/beam and $\simeq90.3\,\mu$Jy/beam respectively.
See Table~\ref{tbl1} for the region definitions.  The rms noise of
each region was estimated from the Gaussian core of the amplitude
distribution of the pixel values as produced by the {\tt AIPS} task
{\tt IMEAN}.  In Figure~\ref{pixhist}, we show the distribution in
pixel values of our entire $33\arcmin$ field. The rms deviation in
peak flux density derived from a fit to this histogram is
$\simeq69\mu$Jy/beam. Figure~\ref{contfull} shows a contour map of our
entire radio field.

Each constant-noise region in Table~\ref{tbl1} was used for the source
extractions. Within each region, we searched for radio sources down to a
peak flux density $\geq5$ times the rms value of the region. The
sources were extracted by the AIPS task {\tt SAD} which uses Gaussian
fits to estimate the fluxes, positions, and angular sizes of the
selected sources.  For faint sources however, unconstrained Gaussian
fits may be unreliable (see Condon 1997). For this reason, we adopted
the following method for the source extraction: first, we ran the task
{\tt SAD} with a $3\sigma_{rms}$ detection threshold to obtain an
initial list of candidates, we then derived the peak flux densities of
the faint sources (with $3\sigma<S_{peak}<7\sigma$) using the {\tt
MAXFIT} task.  
This task uses a second order interpolation algorithm
and is known to be more accurate. 
Only sources with a {\tt MAXFIT}
peak flux density $\geq5\sigma_{rms}$ were retained. For these faint
sources, the total flux density was estimated using the task {\tt
IMEAN}, which integrates the (median background subtracted) 
pixel values in a specific rectangle.
The rectangle was chosen to encompass as much of the source as possible.
For all other parameters (sizes, positions and position angles) we
retained the values obtained from the initial Gaussian fits.  Only two
sources had irregular morphologies showing multiple components.  For
these, the total (background subtracted) 
flux was determined using the task {\tt TVSTAT} which
allows an integration over a specific irregular area defined to
encompass as much of the source as possible.

The numbers of sources detected in each constant-noise region
are summarised in Table~\ref{tbl1}. 
A total of 62 sources were detected with
flux densities $\geq5\sigma_{rms}$ over an area of $0.303\,{\rm deg}^{2}$. 
Within poisson uncertainties,
this number is consistent with source
counts from surveys by Ciliegi et al. (1999) and Hopkins et al. (1998).
To our limiting (mean) sensitivity of 0.35mJy and within $0.303\,{\rm deg}^{2}$, 
they report a source count of typically $70\pm8$.
This confirms the reliability of our radio source detections and
flux density estimates.

\begin{figure}
\plotone{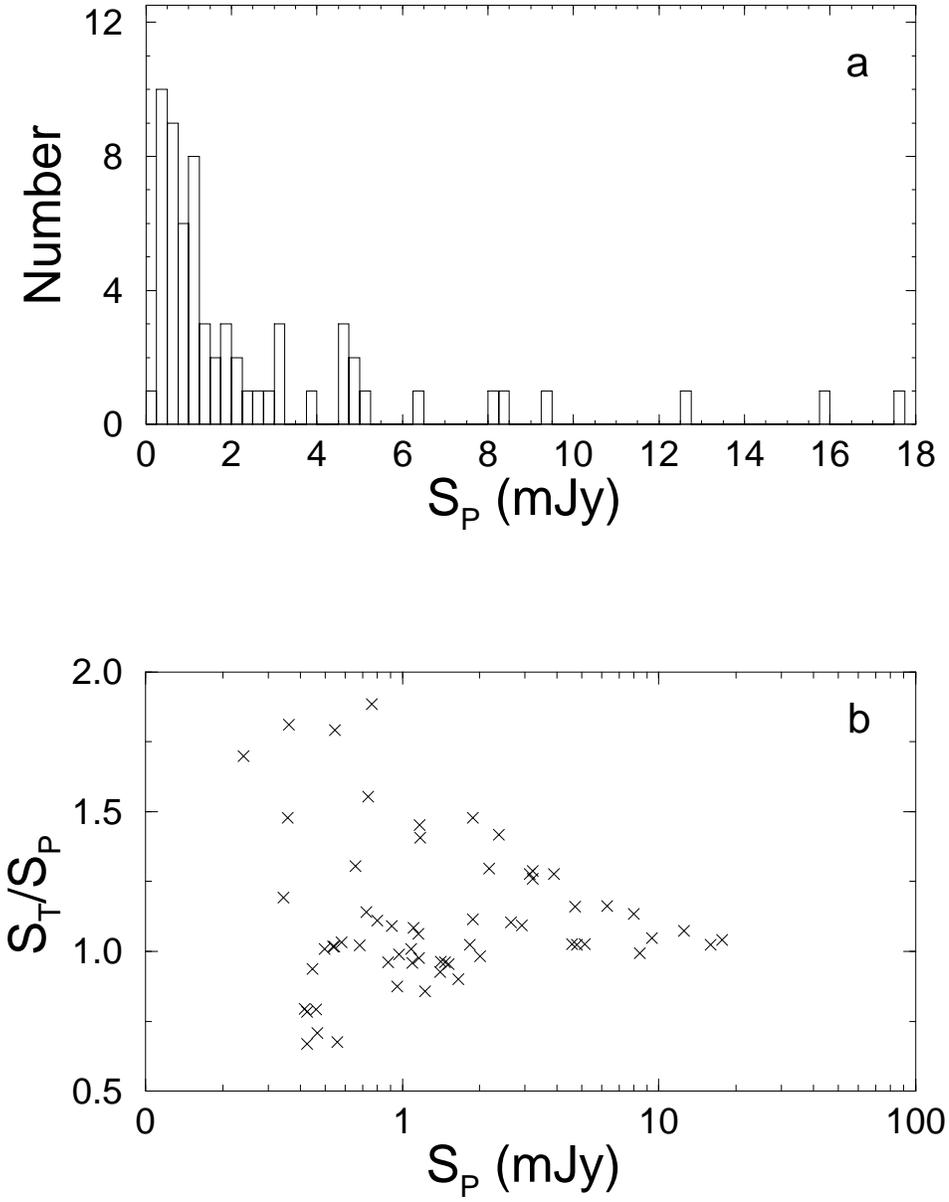}
\vspace{-0.75in}
\caption{(a) Distribution of peak flux density and (b) total
to peak flux ratio as a function of peak flux.
\label{fhist}}
\end{figure}

Table 2 shows the full radio catalog which reports (in column order): 
the source name; RA and Dec.(J2000); errors in RA and Dec.; 
the peak flux density $S_{P}$; error in $S_{P}$; the total flux density
$S_{T}$; error in $S_{T}$; the full width half maximum (FWHM) of the
major and minor axes $\theta_{M}$ and $\theta_{m}$ (determined from
Gaussian fits); 
the position angle
$PA$ of the major axis (in degrees); and the 
rms errors associated with 
$\theta_{M}$, $\theta_{m}$, and $PA$ respectively. 
The different components of multiple
sources are labelled ``C1'' and ``C2''.
In Figure~\ref{fhist}, we show the distribution of peak flux densities and the
total to peak flux ratio as a function of peak flux for all sources. 
Sources with ratios $S_{T}/S_{P}<1$ in Figure~\ref{fhist}b
are primarily due to uncertainties on measured fluxes as estimated from the
two dimensional Gaussian fits. 
Uncertainties in peak fluxes are typically $\lesssim10\%$, while
total flux estimates are more uncertain due to a sensitive dependence
on the Gaussian fitting procedure. Errors in total fluxes are typically  
$\lesssim30\%$. 
Contour maps of radio sources with available 
optical data are shown in 
Figure~\ref{optradim}.

\subsection{Comparison with the {\it NVSS} Catalog}
\label{nvssc}

\begin{figure}
\vspace{-1.5in}
\plotone{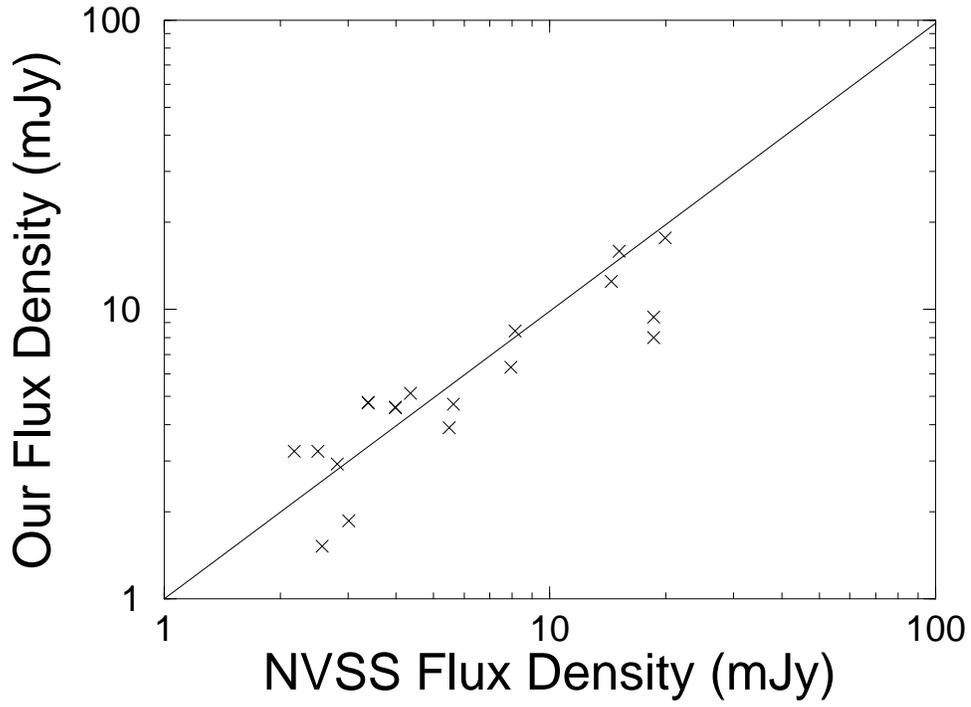}
\caption{The peak flux density obtained with our survey
versus that reported for the same sources in the {\it NVSS} public
database.
\label{nvss}}
\end{figure}

The $33\times33\,{\rm arcmin}^{2}$ region that we observed with the
VLA was also covered by the NRAO VLA Sky Survey (Condon et al. 1998,
NVSS).  The NVSS covers the sky north of $\delta=-40\arcdeg$ at
1.4 GHz with $45\arcsec$ resolution and limiting flux density of
$\simeq2.25$ mJy ($5\sigma$).  To this limit, 17 of our sources were
found to be in common with the NVSS public catalog. 
One source in our catalog however is a double component source and is
unresolved by the NVSS.
A comparison
of flux densities derived in this study with those from this catalog
is shown in Figure~\ref{nvss}.  It is evident that our derived flux
densities are on average slightly lower than those from the NVSS. This
was also reported in a larger comparison study by Ciliegi et
al. (1999) using a similar observational set-up and can be explained
by the difference in resolutions used in the two surveys: $15\arcsec$
vs. $45\arcsec$.  High-resolution surveys
tend to miss flux from low-surface-brightness
emission.  Although the
effect is only marginal for bright sources, it may become important
for attempts to detect faint, low-surface-brightness objects at
high redshift.

\section{OPTICAL PHOTOMETRIC OBSERVATIONS}

\subsection{Observations and Data Reduction}
\label{optobs}

Optical CCD photometry of our radio field was carried out on the
5-m Hale telescope at Palomar\footnote{Operated by California Institute
of Technology, Pasadena, CA 91125} Observatory during the nights of 27-29
August 1997. These fields were initially selected for subsequent
deep mid-IR imaging
with the {\it Wide Field Infrared Explorer} mission ({\it WIRE}),
but the mission failed to perform to expectations.
The Carnegie Observatories Spectroscopic Multislit and Imaging Camera
(COSMIC; Kells et al. 1998) mounted at the
$f/3.5$ prime focus with a TEK 2K CCD was used
to image nine $9.7\times9.7\,{\rm arcmin}^{2}$ fields in the
Gunn $r$ (6550\AA) filter.
Each optical field comprised of three optical pointings
offset by $2\arcmin$, each
with integrations of 600 sec giving a total 1800 sec per field.
This resulted in a limiting magnitude of $r\simeq25$ mag ($5\sigma$).
The seeing was typically $1-1.4\arcsec$ (FWHM).
The optical fields do not cover our entire
$33\times33\,{\rm arcmin}^{2}$ radio field. The
nine slightly overlapping optical
fields correspond to an areal coverage
$27.5\times27.5\,{\rm arcmin}^{2}$, or about 70\% of the radio map.
 
The CCD data were reduced with standard tasks in the {\tt IRAF} package.
Frames were first bias-subtracted, then flat-fielded using dome flats.
Bad pixels and columns were removed by interpolating between adjacent
pixels and lastly, the individual dithered frames were median combined to
remove cosmic ray hits.
Calibration was performed using standard stars from Thuan \& Gunn (1976).
These were used to correct for atmospheric extinction from varying airmass
and provide the instrumental zero point. Photometric uncertainties, estimated
using these standards, are no more than $\approx0.05$ mag.
 
Sources in the reduced optical frames were extracted using the
{\tt DAOPHOT} package in {\tt IRAF} (Stetson 1987). This package has the
benefit of performing photometry in crowded fields, which is the case
in most regions of our optical fields.  It performed the following
steps: first, sources were extracted above a given threshold, given as a
multiple of the total CCD noise (sky and read noise, $\sigma_{tot}$).
We adopted threshold values of 4.5-5$\sigma_{tot}$, just high enough
to avoid large numbers of spurious detections.
Second, simple aperture photometry was performed on these identified sources.
This required a specification of the aperture radius which is likely
to contain most of the light of our target source, and width of a
surrounding annulus to estimate and subtract the sky background. We adopted
a radius of six arcsec and annulus width of four arcsec.
Next, a point spread function (PSF) was
determined in each of our nine fields. This involved an interative
technique to remove contamination from neighbouring sources in
crowded fields near our PSF candidates.
Simultaneous PSF fitting on all
sources was then performed to identify sources which were previously hidden
in crowded regions.
Finally, the magnitudes determined from PSF fits were aperture
corrected to a common aperture size as used on our standard stars. Aperture
corrections were typically 0.22 mag.
A final visual inspection removed any spurious detections. A total of
$\approx300$-390 sources were extracted from each $9.7\times9.7\,{\rm arcmin}^{2}$
field to a limit of $r\simeq25$ mag. Previous optical surveys find
typically 380-520 sources within this area to this limit, and the variation 
is primarily due to clustering. Such fluctuations are found to be 
significant on such
relatively small scales (eg. Metcalfe et al. 1991).

\subsection{Astrometry}
\label{astrom}

The astrometry on the optical images was based on 10-12
APM catlog stars in each field (Maddox et al. 1990). 
The {\tt ccmap} and {\tt cctran} tasks
in the {\tt IRAF} {\tt immcoords} package were used to compute
plate solutions relating pixel positions to astrometric coordinates.
Astrometric coordinates for all sources on the frames were then
determined. By comparing the positions of several sources common to the APM
catalog and our fields, we found that our rms position uncertanties
are typically $\lesssim0.9\arcsec$.

\section{OPTICAL/NEAR-INFRARED IDENTIFICATION OF RADIO SOURCES}
\label{optid}

\subsection{Method for Optical Identification}

We assigned optical identifications and estimate their reliability
using a robust likelihood ratio ($LR$) analysis. This general
method has frequently
been used to assess identification probabilities for radio and infrared sources
(eg. de Ruiter, Willis \& Arp 1977; Prestage \& Peacock 1983; Sutherland \&
Saunders 1992; Lonsdale et al. 1998). The method, which computes the
probability that a suggested identification is the `true' optical
counterpart, is outlined as follows:

For each optical candidate $i$ in the search area of some radio source
$j$, we calculated the value of the dimensionless difference in radio
and optical positions:
\begin{equation}
r_{ij}=\left[\frac{(\alpha_{i}-\alpha_{j})^{2}}{\sigma_{\alpha_{i}}^{2}
+ \sigma_{\alpha_{j}}^{2}} +
\frac{(\delta_{i} - \delta_{j})^{2}}{\sigma_{\delta_{i}}^{2} +
\sigma_{\delta_{j}}^{2}}\right]^{1/2},
\label{rij}
\end{equation}
where the $\alpha$'s and $\delta$'s represent right ascensions and
declinations respectively, and $\sigma$'s standard deviations.
We chose a moderately large search radius of
$10\arcsec$ to allow for the maximal position uncertainties:
$\sigma_{{\rm opt}}\approx1.4\arcsec$ and $\sigma_{{\rm
rad}}\approx1.5\arcsec$ (assuming $5\sigma_{eff}$, where
$\sigma_{eff}=(\sigma_{{\rm opt}}^{2} + \sigma_{{\rm
rad}}^{2})^{1/2}$).  Such a radius is also small enough to avoid large
numbers of chance associations.

Given $r_{ij}$, we must now distinguish between two
mutually exclusive possibilities:
(1) the candidate is a confusing background object that
happens to lie at distance $r_{ij}$ from the radio source or
(2) the candidate is the
`true' identification that appears at distance $r_{ij}$ owing solely to
radio and optical position uncertainties. We assume that
the radio and optical positions would coincide if these uncertainties were
zero. This assumption however is not valid when the centroid of an
extended radio source is used, and is further discussed below.

To distinguish between these cases, we compute the likelihood
ratio $LR_{ij}$,
defined as:
\begin{equation}
LR_{ij}=\frac{\exp\left[-r_{ij}^{2}/2\right]}{2\pi N(<m_{i})
\left[(\sigma_{\alpha_{i}}^{2} + \sigma_{\alpha_{j}}^{2})
(\sigma_{\delta_{i}}^{2} + \sigma_{\delta_{j}}^{2})\right]^{1/2}},
\label{LR}
\end{equation}
where $N(<m_{i})$ is the {\it local} surface density of objects brighter
than candidate $i$.
The likelihood ratio $LR_{ij}$ is simply the ratio of the
probability of an identification (the Rayleigh distribution:
$r\exp{(-r^{2}/2)}$),
to that of a chance association at $r$ ($2\pi N(<m_{i})\sigma_{\alpha}
\sigma_{\delta}$). $LR_{ij}$ therefore represents a `relative weight'
for each match $ij$, and our aim is to find an optimum cutoff $LR_{c}$
above which a source is taken to be a reliable and likely candidate.
Its advantage over alternative methods (purely based on
finding the lowest random chance match, eg. Downes et al. 1986) 
is that it allows for
a possible distant candidate to still be the `true' identification
even when there is still a high chance of it being a spurious
background source.

It is important to note that our form for $LR_{ij}$ (eq.~\ref{LR})  
slightly differs from that used by earlier studies (eg. Lonsdale et al. 1998) 
in that it doesn't contain the multiplicative ``$Q$'' factor in the
numerator. This factor represents the apriori probability that a ``true''
optical counterpart brighter than the flux limit exists amongst the
identifications. For our purposes, we will treat $LR_{ij}$ as simply a
a relative weight measure for each radio-optical match, just for the
purposes of assigning an optimal cutoff for reliable
identification (see below). We are not concerned
with its absolute value, which is required when computing formal
probability measures from $LR$. For simplicity, we have therefore set
$Q=1$ in this work.

The optical surface density as a function of magnitude
to be used in computing $LR$ was
determined from the total number of objects visible
in our optical frames.
The variation in surface density
in the vicinity of
each radio source caused by possible clustering effects was found to be small:
no more than 5\% on $2-3\arcmin$ scales.

\begin{figure}
\plotone{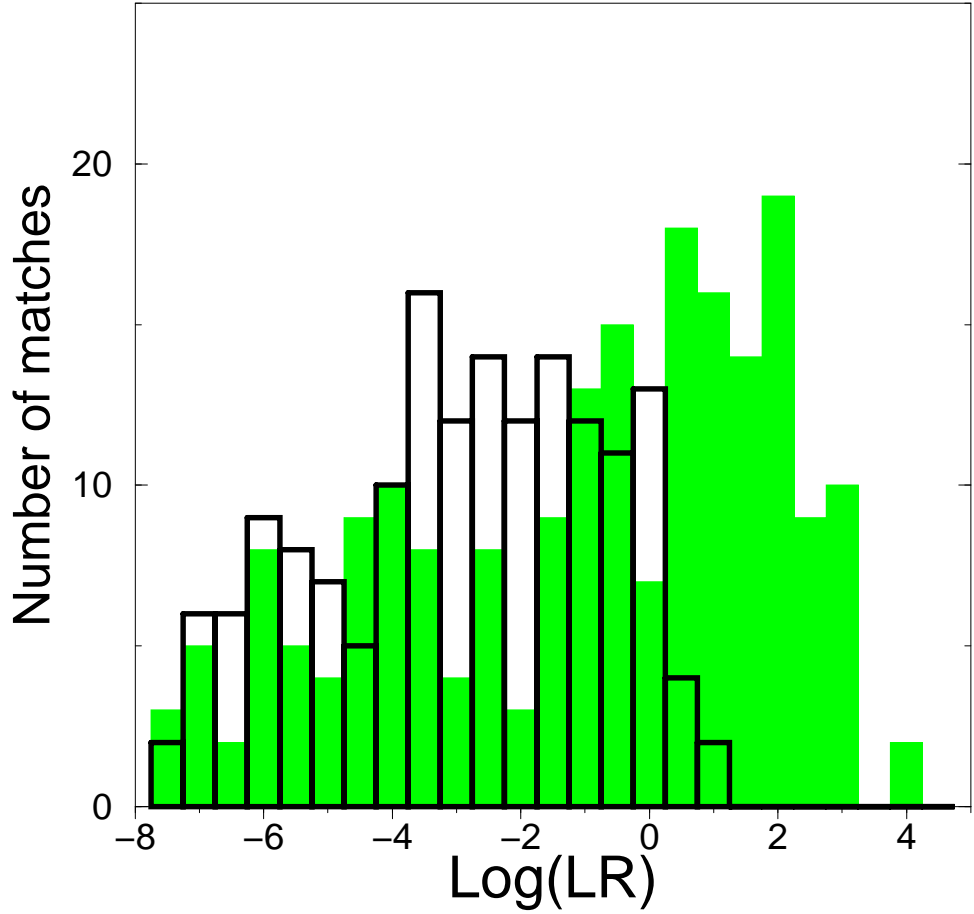}
\vspace{-1.5in}
\caption{Distribution of $LR$ for radio-optical matches at 
``true'' radio positions
(shaded histogram), and at ``random'' radio positions (thick-lined histogram). 
\label{LRhisto}}
\end{figure}

The distribution of $LR$ values for all possible radio source-candidate matches 
is shown by the shaded histogram in Figure~\ref{LRhisto}. 
Following Lonsdale et al. (1998), we generate a truly
random background population with respect to the radio sources by 
offseting the radio
source positions by $\approx30\arcsec$. $LR$ values for each
radio source were then re-computed and their distribution is given by
the thick-lined histogram in Figure~\ref{LRhisto}. 
A comparison of the number of associations for (true) radio source
positions with the number of associations found for random (offset) positions
will enable us to determine a critical value $LR_{c}$ 
for reliable identification.
From these distributions, we compute the reliability
as a function of $LR$:
\begin{equation}
R(LR_{ij})=1-\frac{N_{random}(LR_{ij})}{N_{true}(LR_{ij})},
\label{reliability}
\end{equation}
where $N_{true}$ and $N_{random}$ are the number of true and random
associations respectively.
The reliability computed in this way
also represents an approximate measure of the identification
probability for a candidate with given $LR$.

\begin{figure}
\plotone{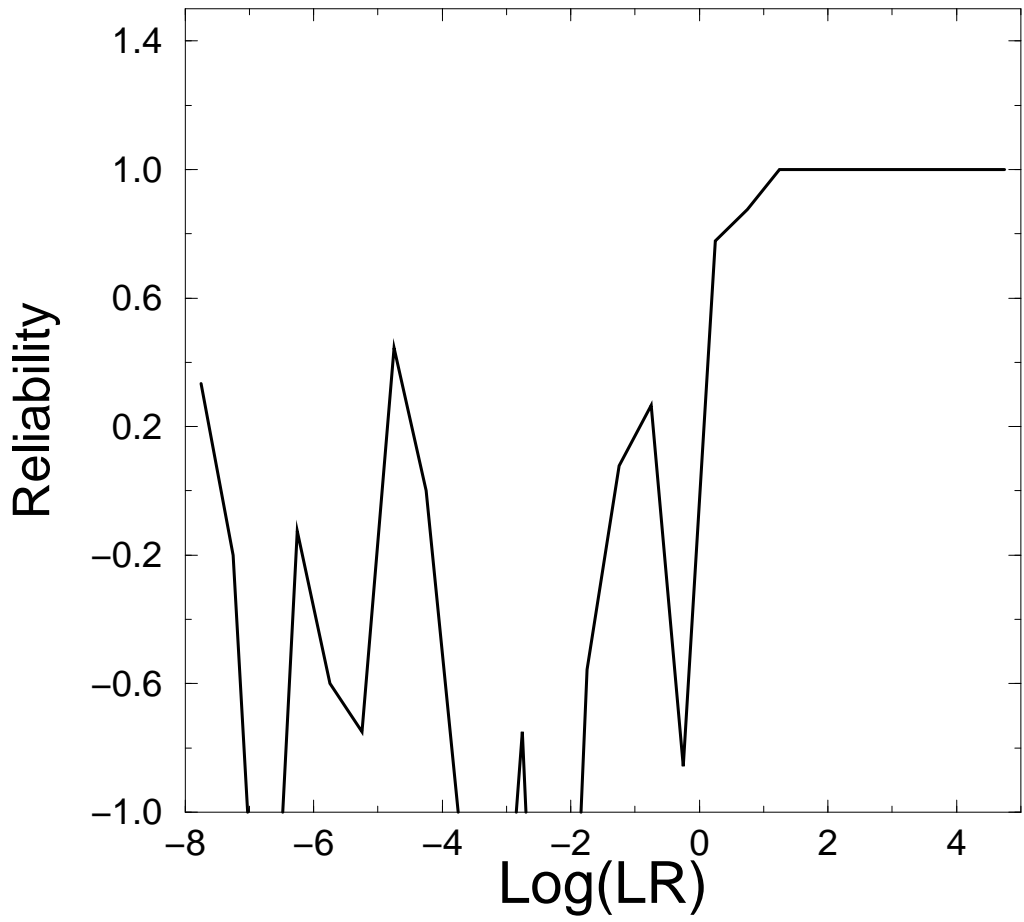}
\vspace{-1.5in}
\caption{Reliability as a function of $LR$. See equation
(\ref{reliability}).
\label{relfig}}
\end{figure}

Figure~\ref{relfig} illustrates the reliability as a function of $LR$. 
Above $\log{(LR)}\sim0.5$, the reliabilities are $\gtrsim85\%$ because
few random associations exceed this value of $LR$ (Fig.~\ref{LRhisto}). 
As a good working measure
we therefore assume a cutoff $\log{(LR_{c})}\simeq0.5$ above which
a source is taken to be a likely candidate.
The determination of reliabilities via the 
$LR$ method is insensitive to variations in $N(<m_{i})$ across the
field or uncertainties in its derivation, and also the assumption of
Gaussian error ellipses in the radio and optical positions.
Such uncertainties are ``normalised out'' when one computes the ratio 
of random to true number of associations within a search radius
when estimating the reliability (eq.~\ref{reliability}).
Lonsdale et al. (1998) have shown that the absolute value of $LR$ itself
depends on the characteristics of the source population being identified
(eg. stars versus galaxies).
Different populations (assuming they could be classified 
a-priori using some diagnostic) map into different underlying surface
densities at the `identifying' wavelength, implying that
distributions in $LR$ (eq.~\ref{LR}) will also be different. 
For a robust determination of the reliability in such situations, see
Lonsdale et al. (1998).

There are two complications to consider in the above method. 
The first is
when one attempts to identify extended (or resolved)
radio sources with this method. 
For all radio sources, we have used the positions of centroids 
derived from two-dimensional Gaussian fits in computing the $LR$
for optical candidates. 
For unresolved sources with (Gaussian fitted) sizes
$\lesssim15\arcsec$ (the synthesised FWHP beam width), the source is likely
to have a compact central component and the
optical position is expected to lie close to its quoted radio centroid. 
For an extended (resolved) source however, the 
radio and optical positions may differ considerably since errors in the
radio centroids are only $\lesssim2\arcsec$. 
In such cases, if $LR<LR_{c}$ the
identification may still be valid, since its low $LR$ value could
purely be due to a real large positional offset.
The second complication is when a radio source has more than one
optical candidate within its search radius with $LR>LR_{c}$. 
This occured in about 20\% of cases and was primarily due to
contaminating stars. 
We assess these ambiguities and increase the robustness of our
identifications by visually examining all optical candidates according
to the following criteria:  
\begin{enumerate}
\item{If candidates have $LR<LR_{c}$ for a radio source with 
$\theta_{min,max}\gtrsim15\arcsec$,
then identification is classified as {\it uncertain}.}
\item{If candidates have {\it very low} 
reliability, $LR\ll LR_{c}$ (for {\it unresolved} radio sources), 
or there are no objects in
the search radius, then
radio source is classified as {\it empty field}.}
\item{If $LR\lesssim LR_{c}$, i.e. where reliability is 
{\it moderately ``low''}, then identification is also uncertain.}
\item{If more than one optical candidate exists with $LR>LR_{c}$, then only
source(s) with extended (galaxy-like) 
optical profile is taken as the identification. Point sources associated
with quasar nuclei are not considered in our identification scheme
due to their relatively low
surface density compared to galaxies ($\approx 1:4000$) 
in sub-millijansky
radio samples.} 
\item{For unique, $LR>LR_{c}$ candidates, its optical profile is also
checked for confirmation.}
\end{enumerate}

\subsection{Results}
 
Of the 62 radio sources, 43 lie within our optically imaged
$27.5\times27.5\,{\rm arcmin}^{2}$ field.  We found optical
identifications for 26 to $r\simeq25$ mag with reliabilities
$R_{id}\gtrsim80\%$. 
Four sources have identifications
classified as uncertain owing to a moderately low identification
reliability of $R_{id}\lesssim78\%$ (and
$\log{(LR)}\lesssim0.4$). Five more are uncertain because they have
extended radio structure and large possible positional offsets between
optical and radio centroid positions. Eight radio sources lie in
``definite'' optical empty fields with no candidates
brighter than $r\simeq25$ mag.

\begin{figure}
\plotone{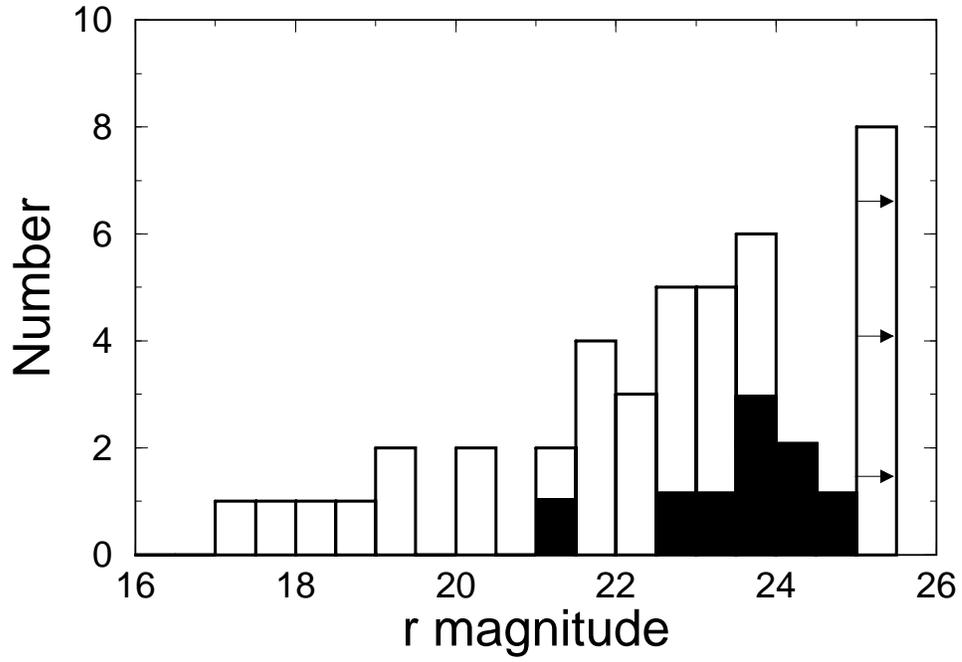}
\vspace{-1.5in}
\caption{The distribution of apparent $r$-band magnitudes
for the 26 identified radio sources (open histogram) and nine uncertain
identifications (filled histogram). Eight empty fields are 
represented by the $r>25$ magnitude bin with arrows.
\label{rhisto}}
\end{figure}

Other optical follow-up studies found similar results.  Georgakakis et
al. (1999) identified $\approx47\%$ of sources to $R=22.5$ mag from
the Phoenix Deep Survey ($S_{1.4{\rm GHz}}>0.2$ mJy) (Hopkins et
al. 1998). Deeper indentifications of sources as faint as $S_{1.4{\rm
GHz}}\simeq40\mu$Jy from Hubble Deep Field images revealed a 80\%
success rate to $I=25$ mag (Richards et al. 1999).  Ignoring the
uncertain identifications in our study (from criteria 1 and 3 above and
which are excluded from our analysis),
we find that $\approx$18\% of our sources are
unidentified to $r\sim25$.  Accounting for differences in bandpasses
and sensitivity, this is broadly consistent with the above studies.
Figure~\ref{rhisto} shows the distribution of apparent magnitude $r$
for all reliable (robust) and uncertain identifications in our sample.
 
\subsection{Near-Infrared Data}

Near-infrared data in the $J$(1.25$\mu$m), $H$(1.65$\mu$m) and
$K_{s}$(2.17$\mu$m) photometric bandpasses were obtained from the Two
Micron All Sky Survey ({\it 2MASS}) project database.  For multi-band
detection of point sources, this survey is currently scanning the sky
to sensitivities 16.5, 16.0 and 15.5 mag at signal-to-noise ratios
$\approx7$, $\approx5$ and $\approx7$ respectively in $J$, $H$ and
$K_{s}$.  
The data relevant to this study are not yet released in the public catalogs, 
and was retrieved from the `internal working database' at {\it
IPAC}\footnote{The Infrared Processing and Analysis Center, California
Institute of Technology.}.
Photometry in this database was determined using custom PSF-fitting
software and algorithms are described in Cutri et al. (2000). 
Since such data have not been subjected to the
rigorous quality assurance as that in the public release catalogs, we have
examined individual images for quality and any possible systematic 
uncertainties in the photometry.

To maximise the possible number of detections, we 
searched for near-infrared counterparts to each radio source
with a conservatively low signal-to-noise ratio threshold of $\sim5$
in each band.
In cases where a source was detected in only one or two of the
three bandpasses: $J$, $H$ and $K_{s}$, we note its ``band-filled'' 95\%
confidence upper-limit in the undetected band. In other words, the 
{\it 2MASS}
catalog also reports upper limits to measurements in an undetected band
by placing an aperture over the position inferred from detections in other
bands.
The image pixel scale of the {\it 2MASS} detectors
is $2.0\arcsec$ and the positional uncertainties are
$\lesssim0.5\arcsec$.
Owing to the relatively shallow flux limits of the {\it 2MASS} survey, the
background source surface density is low enough that
chance associations with radio positions are very unlikely.

We searched the {\it 2MASS} database for near-infrared counterparts to our
43 radio sources that have {\it available optical} information 
from our deep optical survey and examined 
their images for quality.
We found 7 reliable matches with 6 detections in
$J$, 7 in $H$, and 4 in $K_{s}$ at
the $\gtrsim5\sigma$ level. Upper limits were available for
the remainding ``band-filled'' values.

Results of our optical identification analysis and 
available near-infrared data are shown in Table 3. 
In column order, this table reports: the radio source name 
(see Table 2); RA and Dec.(J2000) of the optical counterpart
of the radio source; optical-radio position separation ($\delta_{rad-opt}$)
in arcsec;
logarithm of the liklihood ratio ($LR$); reliability of the optical
identification (see eq.[\ref{reliability}]);
apparent $r$-band magnitude; $J$, $H$ and $K_{s}$ magnitudes with
errors or $2\sigma$ upper limits; $r-H$ color; $r-K_{s}$
color, and last, the optical morphology if the optical counterpart is
visually resolved with size $\gtrsim5\theta_{FWHM}(PSF)$. 

Optical morphologies were determined from light
profile fitting of individual sources and 
comparison with $R^{1/4}$-law (elliptical-like)
and exponential (disk-like) profiles. In cases where a disturbed or
interacting morphology is apparent, then it is designated to have
an irregular (labelled as `Irr') 
morphology. Sources unresolved in the optical with
typically $\lesssim5\theta_{FWHM}(PSF)$ are designated as `unknown' and
labelled as `?' in Table 2. 

\begin{figure}
\vspace{-1in}
\plotone{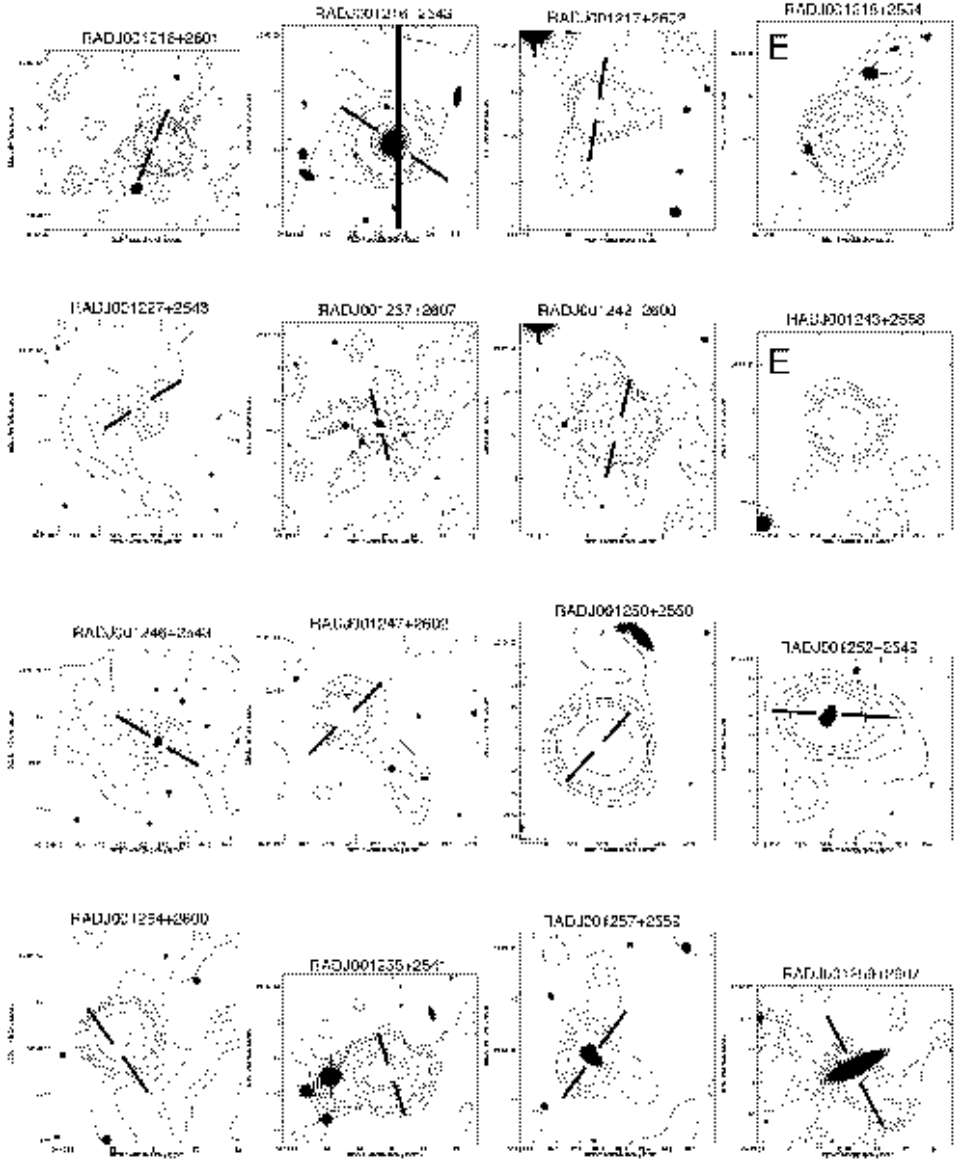}
\vspace{-1.2in}
\caption{Optical image ($r$-band) and radio contour overlays for the
43 sources with available optical information.
Contour levels are 
$1\sigma$, $2\sigma$, $3\sigma$, $4\sigma$, $5\sigma$, 
$10\sigma$, $20\sigma$, $30\sigma$, $40\sigma$, $50\sigma$, 
$100\sigma$ and $200\sigma$,
where $\sigma$ refers to the local rms noise (see Table~\ref{tbl1}).
Optical candidates are indicated within broken lines---see Table 3 for
reliability estimates. Maps labelled with ``E'' in upper left corner
represent optical empty fields. 
\label{optradim}}
\end{figure}

\begin{figure*}
\vspace{-1in}
\plotone{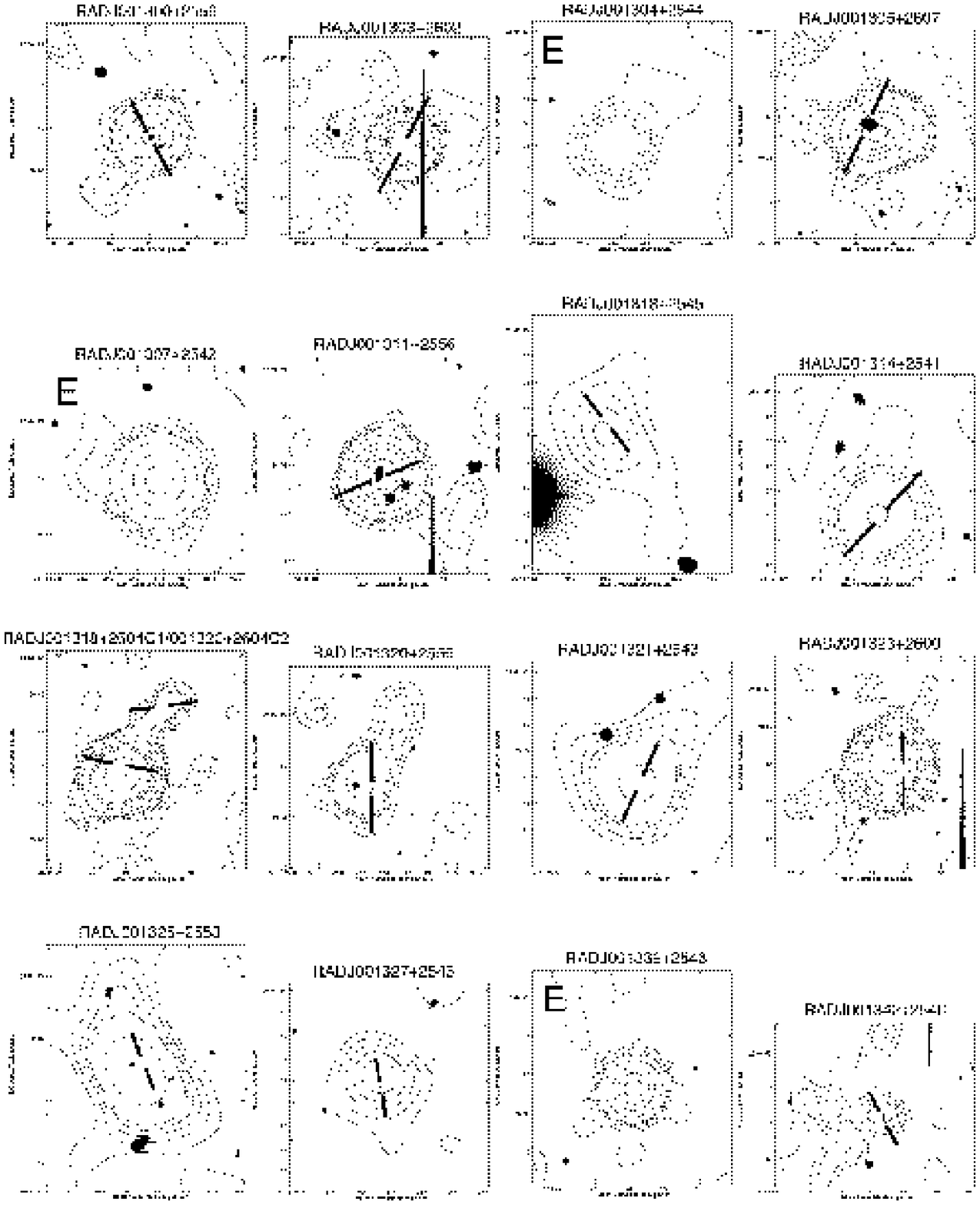}

Fig. 8.--- contd...

\end{figure*}

\begin{figure*}
\vspace{-1in}
\plotone{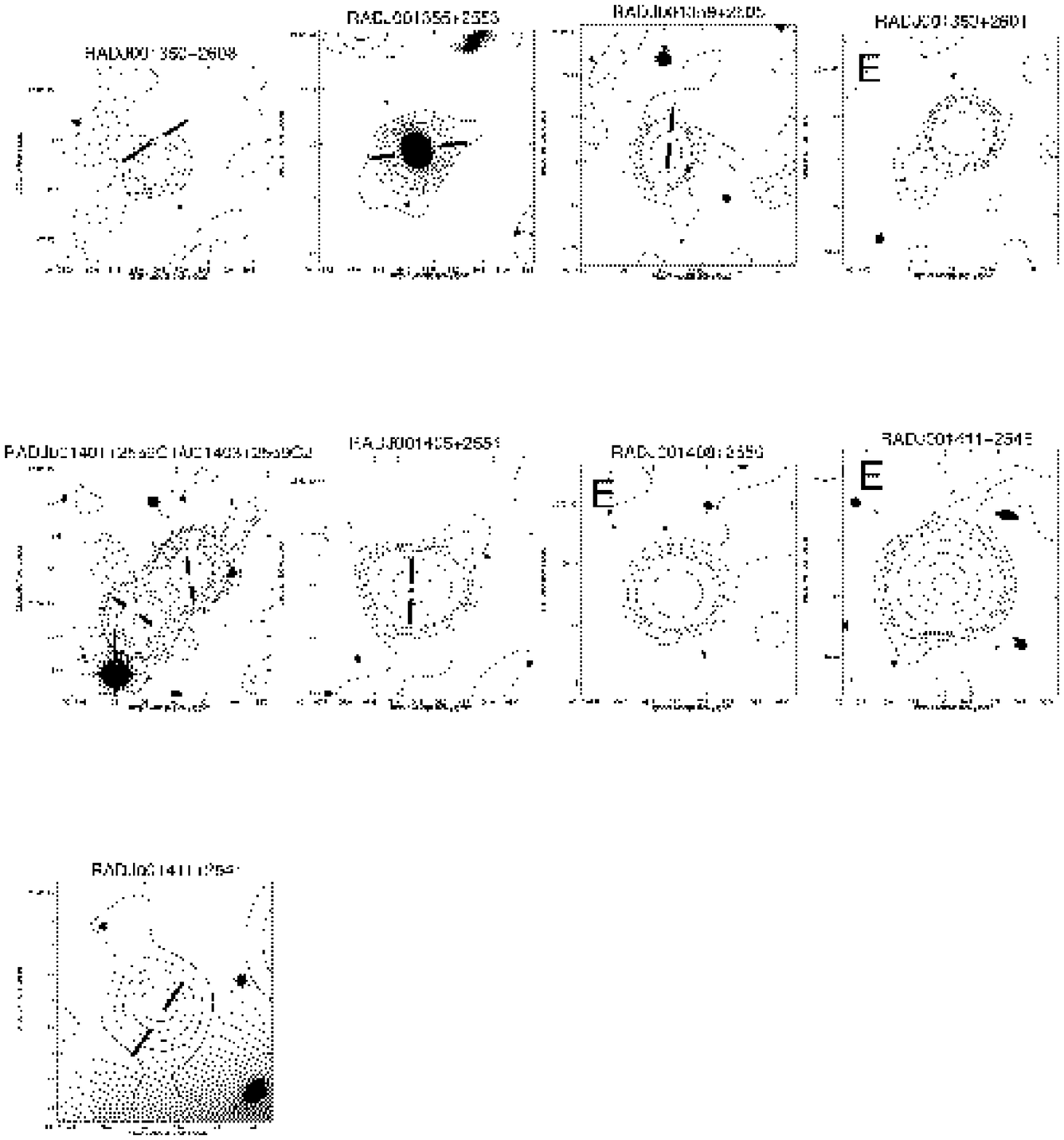}

Fig. 8.--- contd...

\end{figure*}

\subsection{Optical and Radio Map Overlays} 
\label{optidrad}

In Figure~\ref{optradim} we show the optical image -- radio map overlays
for the 43 radio sources with available optical information. A visual
inspection of the optical images of resolved counterparts shows a
diverse morphological mix, consistent with previous studies.
About 40\% of our optical identifications have elliptical/disk-like morphology,
while $\sim10\%$ can be identified as exhibiting peculiar
(either interacting or disturbed) morphologies.
It is important to note that these `disturbed' sources are based on
visual inspection alone and their morphology could still be uncertain 
until future spectroscopy confirms their nature. 
The elliptical/disk hosts also tend to be associated on average with
sources of relatively
brighter radio flux density ($\gtrsim2$ mJy) than the irregular
class. This is consistent with previous radio-optical identification
studies which find an increasing fraction of irregular-type galaxies
at $S_{1.4}\lesssim 2$ mJy (eg. Kron, Koo, \& Windhorst 1985,
Hammer et al. 1995, Gruppioni et al. 1998) and
a significant number of
elliptical galaxies hosting the brighter
extended radio galaxies and AGN (Condon 1989).
 
A further observation is the unique
radio structure exhibited by our eight optical empty field sources with
$r>25$ mag. 
These are represented by maps in Fig.~\ref{optradim} labelled by the letter
``E''. All show
compact and symmetric (presumably unresolved) structures
and could represent either of the following:
distant (possibly dusty) AGN where with our radio sensitivity,
we could have detected a nominal FR-I galaxy to $z\sim1.3$, or,
nearby, compact dusty starbursts at $z\lesssim0.3$ as constrained
by typical starburst luminosities: 
$L_{1.4{\rm GHz}}\lesssim10^{23}{\rm W}\,{\rm Hz}^{-1}$.
The second explanation for the nature of the empty fields
is more plausible, given that the majority of submillijansky radio
sources are associated with star-forming galaxies and less than 5\%
are usually identified with bright FR-Is at $z\lesssim1$
(Kron, Koo \& Windhorst 1985).

\placefigure{optradim}

\section{ANALYSIS OF RADIO AND OPTICAL--NEAR-INFRARED COLORS}

This section presents an analysis of flux ratios between the radio,
near-infrared, and optical bands to explore possible contributions
from AGN and starbursts to the observed radio emission as well as the
importance of absorption 
by dust.  Because we lack spectroscopic
information, our analysis treats the sub-mJy sources as one
homogeneous population and uses a simple stellar synthesis model to
interpret its properties quantitatively.

\subsection{A Simple Synthesis Model} 
\label{profr}

We can predict the radio--to--optical(--near-IR) flux ratios and 
$r-K$ colors for a range of galaxy types using the stellar population
synthesis code of Bruzual \& Charlot (1993) (hereafter BC). On its own 
however, the BC model does not directly predict the amount of
radio emission expected from a star-forming galaxy, which we need
for the determination of flux ratios involving the radio band.
We do this by relating the star-formation rates derived from
empirical calibrations involving the UV and radio bands as follows. 

The 1.4 GHz radio emission from star-forming systems is believed to
be primarily synchrotron emission from cosmic rays accelerated 
in supernova remnants plus
a small ($\sim10\%$) thermal contribution from H{\small II} regions 
(Condon \& Yin 1990, Condon 1992). Thus to a good approximation,
the radio luminosity is taken to be proportional to the formation rate
of stars with $M>5M_{\sun}$:
\begin{equation}
SFR(M\geq5M_{\sun})=\frac{L_{1.4{\rm GHz}}}{4\times10^{28}{\rm
erg}~{\rm s}^{-1} {\rm Hz}^{-1}}M_{\sun}{\rm yr}^{-1}.
\label{SFRrad}
\end{equation}
(Condon 1992).
These same massive stars will also contribute significantly to the
UV continuum emission in the range $\sim1200$-$2500$\AA. In particular,
there have been many different calibrations of the SFR from the UV-flux. 
For a Salpeter initial mass function (IMF) from $mM_{\sun}$ to 100$M_{\sun}$,
the calibration of Madau et al. (1998) (which assumes {\it no dust}
correction) yields 
\begin{equation}
SFR(M\geq mM_{\sun})=Q_{m}\left(\frac{L_{UV}}{7.14\times10^{27}{\rm
erg}~{\rm s}^{-1}{\rm Hz}^{-1}}\right)M_{\sun}{\rm yr}^{-1}.
\label{SFRUV}
\end{equation}
We have modified the initial relation of Madau et al. (1998) to 
include the factor $Q_{m}$, which represents the fraction of stellar masses 
contributing to the observed SFR,
\begin{equation} 
Q_{m}=\frac{\int^{100M_{\sun}}_{mM_{\sun}}M\psi (M)dM}
{\int^{100M_{\sun}}_{0.1M_{\sun}}M\psi (M)dM}, 
\label{Q}
\end{equation} 
where $\psi (M)\propto M^{-x}$ is the IMF. For $m=0.1$, we have $Q_{m}=1$.  
Assuming a Salpeter IMF ($x=2.35$), we find that
for stars with $M>5M_{\sun}$, $Q_{m}\simeq0.18$. 
With this fraction, and 
equating the two SFR calibrations (eqs [\ref{SFRrad}] and [\ref{SFRUV}]), we
find that the luminosity densities at 1.4 GHz ($L_{1.4GHz}$) and
$\simeq2100$\AA~ ($L_{UV}$) are very nearly equal.  
We therefore assume that the rest frame 1.4 GHz flux density is directly 
given by the flux density at $\simeq2100$\AA~ as specified by the
synthesized model spectrum.
In general terms, the observed radio flux (in the same units as the 
synthesised
UV spectrum) can be written:
\begin{equation}
f_{\nu}({\rm 1.4GHz})_{obs} = (1+z)^{1-\alpha}f_{\nu}({\rm 2100\AA})_{rest}, 
\label{radfl} 
\end{equation}
where $\alpha\simeq0.8$ is the radio spectral index (Condon 1992) and
$f_{\nu}({\rm 2100\AA})_{rest}$ is the rest frame (unreddened) 
UV spectral flux. 
We must emphasise that this relative
radio flux is only that associated with 
the star-formation
process. Possible additional contributions, such as contaminating AGN,
are not considered in this model.

We calculated
flux ratios involving the radio, near-infrared and optical bands 
using
evolutionary synthesis models
for ellipticals (E/SO), early (Sab/Sbc) and late (Scd/Sdm) type spirals, and
``very blue'' starbursts (SB). 
These are meant to represent the possible contributions to
the sub-mJy radio sources, and each class is defined by a characteristic
star formation rate as a function of time. 
As supported by local observations (eg. Gavazzi \& Scodeggio 1996), 
we assumed that E/SO and Sab/Sbc galaxies
have an exponentially decaying SFR of the form 
$\psi(t)\propto\tau^{-1}\exp{(-t/\tau)}$, where $\tau$ is the e-folding time. 
Values of $\tau=1$ and $\tau=8$ Gyr were adopted for the E/SO and Sab/Sbc
galaxies respectively. For late-type spirals (Scd/Sdm) and young
starbursts (SB), we assumed constant SFRs with different ages. 
All models used to generate the spectral energy
distributions (SEDs) are summarised in Table~\ref{tbl4}.
The models assume ${\rm H}_{0}=50\, \rm km\,s^{-1}\,Mpc^{-1}$, 
$q_{0}=0.5$ and a galaxy formation redshift $z_{f}=10$, 
which corresponds to an age of 12.7 Gyr at $z=0$. 
 
To explore the effects of dust on our flux ratios and colors, each model SED  
was reddened 
in the source rest frame with an extinction curve 
$\xi(\lambda)\equiv A_{\lambda}/A_{V}$ characteristic of the Small
Magellanic Cloud (SMC), which appears to be a good approximation for
the ISM of nearby galaxies (Calzetti et al. 1994). 
This approximation is most accurate for the reddest wavelengths 
of starburst galaxy spectra ($7000$\AA-$3\mu$m), although breaks down
severely at $\lambda\lesssim2500$\AA (Calzetti et al. 1999).
We have used the analytical fit for $\xi(\lambda)$ 
as derived by Pei (1992) for the SMC. For simplicity, we assumed that the dust
is distributed in a homogeneous foreground screen at the source redshift. 
 
\subsection{Data and Model Comparions}
\label{comp}

\begin{figure}
\plotone{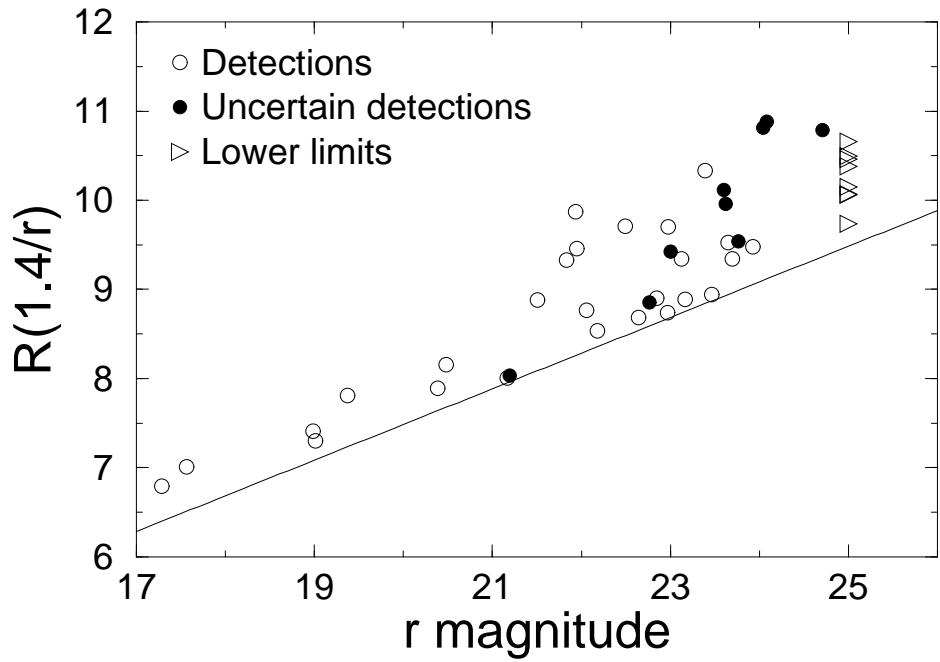}
\vspace{-1.5in}
\caption{Radio--to--optical ($r$-band) flux ratio 
(see eq.[~\ref{RO}]) as a function of $r$ magnitude for all radio sources
with available optical information. 
For the observed range of r-magnitudes, the solid line represents the
prediction at the limiting flux of the survey
$S_{1.4{\rm GHz}}\approx0.3$ mJy. 
\label{Rvsr}}
\end{figure}

In Figure~\ref{Rvsr} we plot the radio--to--optical flux ratio, $R(1.4/r)$,
defined as
\begin{equation}
R(1.4/r)=\log{(S_{{\rm 1.4}}/{\rm mJy})} + 0.4\,r, 
\label{RO}
\end{equation}
where $S_{1.4}$ and $r$ are the radio flux and optical $r$-band
magnitude respectively.
The distribution seen in observed values of $R(1.4/r)$ is larger at the
faintest optical magnitudes $r>21.5$. 
There are few galaxies however with $r<21.5$, and nonetheless,
the scatter at $r>21.5$ 
is consistent with that found at $r\lesssim19$ 
in a larger follow-up 
study of sub-mJy radio sources by Georgakakis et al. (1999). 
 
\begin{figure}
\plotone{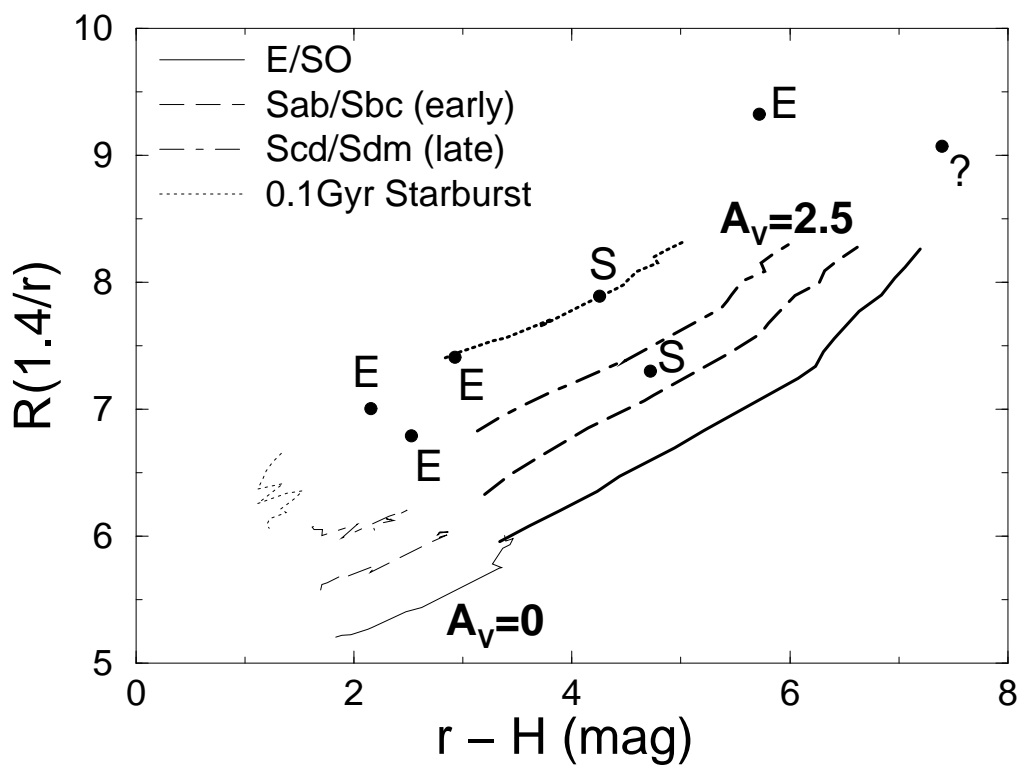}
\vspace{-1.7in}
\caption{Radio--to--optical ($r$-band) flux ratio 
(see eq.[~\ref{RO}]) as a function of $r-H$ color for radio sources
with available optical and near-infrared detections. 
The predictions of
four synthesis models for $0\leq z\leq 1.5$ 
($z=0$--starting bottom left of curves to
$z=1.5$--top right) are also shown. 
These assume no
extinction (thin curves) and with a rest-frame extinction $A_{V}=2.5$ 
mag (thick curves).
\label{Rr}}
\end{figure}

Figure~\ref{Rr} shows $R(1.4/r)$ as a function of $r-H$ color
for all sources with available optical and near-infrared data.
Our reason for using $r-H$ color is that the $H$ band yielded more
``definite'' detections 
than the other near-infrared bandpasses. Although the numbers
are still relatively small,
this facilitates the best comparison with the synthesis models.
The predictions for four galaxy types 
(see Table~\ref{tbl4} and \S 5.1) are shown for no dust reddening (thin curves) and 
a rest-frame extinction $A_{V}=2.5$ mag.
The morphological mix of data shows a relatively large
dispersion in $r-H$ color that is more consistent 
with the range predicted by the models that {\it include dust}. 
This
suggests that on average,
the optical--to--near-infrared continua of most 
sources in Fig.~\ref{Rr} are reddened by a uniform 
(possibly ``optically thin'')
dust component with $A_{V}\simeq2-2.5$ mag absorption. 
This measure is consistent with spectroscopic 
studies of nearby starbursts by Calzetti et al. (1996), Meurer et al. (1997)
and photometric modelling by Nakata et al. (1999). 

We must emphasise that our models only account for radio emission
produced from star-formation processes. The sources labelled as
elliptical (or early type) in Fig.~\ref{Rr} are not expected to lie
on any of the star-formation derived locii. An AGN most likely
dominates their radio emission. We include them here merely as 
a comparision, and their relationship to normal starbursts is discussed
further below.

\begin{figure}
\plotone{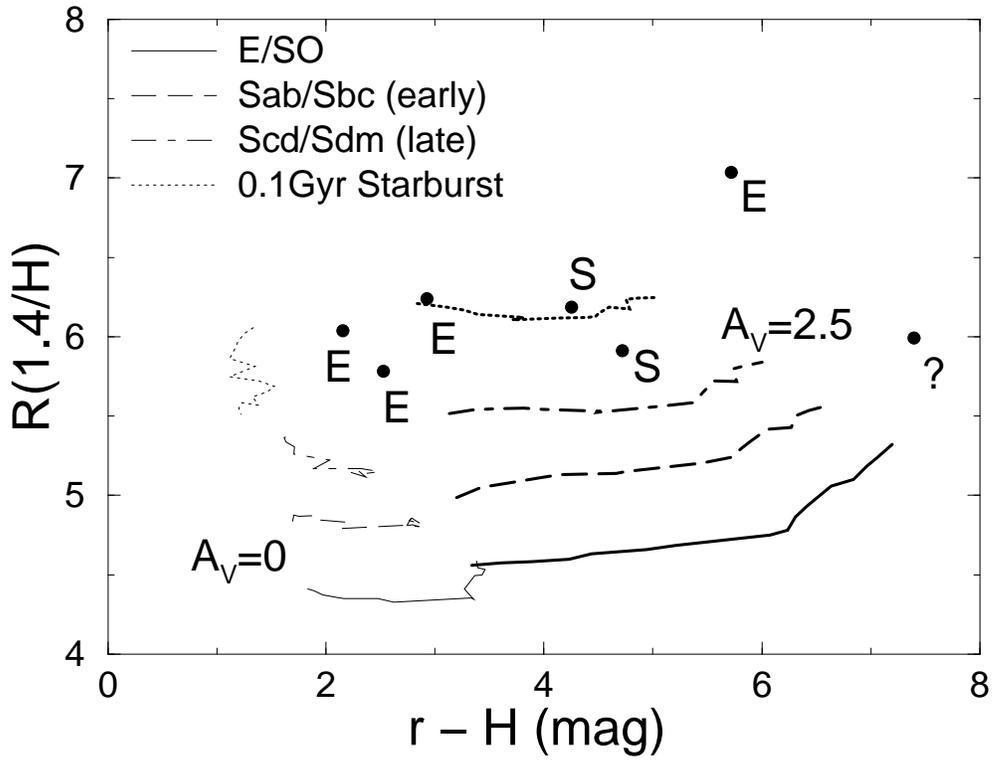}
\vspace{-1.7in}
\caption{Radio--to--near-infrared ($H$-band) flux ratio as
a function of $r-H$ color and models which include no extinction (thin curves)
and $A_{V}=2.5$ mag (thick curves).
\label{Rk}}
\end{figure}

The sources in Fig.~\ref{Rr} appear more-or-less consistent with
the dusty ``0.1 Gyr starburst'' model. This could in principle apply to the 
two sources
with spiral/disk-like morphology (labelled ``S''), but is unconventional for
the five elliptical morphologies (see below). 
A comparison with radio--to--near-infrared flux ratios  
further constrains the underlying properties of these sources.
Figure~\ref{Rk} shows the radio--to--near-infrared flux ratio $R(1.4/H)$,
defined
analagous to eq.[\ref{RO}], as a function of $r-H$ color. The near-infrared
emission is dominated by old stars and is less affected by dust than
the optical. The radio--to--near-infrared flux ratio should therefore
be relatively insensitive to dust.
Given the simplicity of our models, 
the two disk-like sources may not necessarily represent
``0.1 Gyr starbursts'' as indicated in Fig.~\ref{Rk}. 
They could also belong to the Sab/Sbc or Scd/Sdm
classes. For this to be true, the following additional components may play
an important role in more `realistic' models:
`optically-thick' dust that {\it completely obscures}
both the $H$ and $r$ 
emission without causing appreciable reddening in $r-H$ color, or, 
contamination by at least an order of magnitude times more radio emission
from a central AGN than that produced purely by supernovae. 
The second possibility
is favored by radio observations of a number of luminous infrared galaxies
by Norris et al. (1988), where some showed evidence for
significant radio emission from compact Seyfert-like nuclei. 
 
The large discrepancy between the four sources with elliptical morphologies 
(labelled ``E'')
and predictions from the early-type E/SO models suggests the importance
of a significant AGN contribution to the radio emission. 
Appreciable amounts of
optically-thick dust suppressing the optical and near-infrared 
light (except for
extinction by diffuse, optically-thin dust) is not favored by observations
(eg. Goudfrooij \& de Jong 1995).
Most, if not all of these ellipticals are likely to be 
radio-powered by AGN.
At the limiting sensitivity of our radio survey ($\simeq0.3$ mJy), 
a nominal FR-I galaxy (with $L_{1.4}=10^{24}\,{\rm W}{\rm Hz}^{-1}$) could 
be detected
to $z\simeq1.3$, and indeed the spread in $r-H$ color for the ellipticals
in Fig.~\ref{Rk} is consistent with the E/SO (optically-thin dust) model to
this redshift. 
A comparison between the model $R(1.4/H)$ values with actual observed values 
implies that such AGN will contribute a factor 
$10^{2}$ times more radio emission than that produced 
by any underlying star formation activity in these systems.
It is important to note that the ratio of AGN-to-stellar powered
radio activity has a huge spread for the elliptical population in general,
and that the factor $10^{2}$ only illustrates a property
specific to the ellipticals in our radio sample.
 
To summarise, our use of a simple synthesis model that includes radio
emission and dust reddening to analyse the properties 
of sub-mJy radio sources has shown the following: 
first, the presence of dust with extinctions $A_{V}\simeq2$ mag 
and possibly greater, consistent with previous more
direct determinations, and second, that the level of 
radio emission from non-stellar
processes such as AGN could be easily inferred and constrained. 
This will be particularly important for starbursts hosting Seyfert
nuclei where a comparison with more sophisticated dust models
may be required to infer the relative contributions. 
 
\section{A Method to Select ``ULIGS'' via Radio/Optical Color}

Since the emission (and dust absorption) 
properties from normal galaxy populations are reasonably
well known, a color-color diagram such as Fig.~\ref{Rr} 
could provide a potential diagnostic for selecting 
ultraluminous infrared galaxies (ULIGS) to high redshift. 
The relatively low 
sensitivity of the Infrared-Astronomical Satellite (IRAS)
has primarily confined ULIG selection to the local Universe 
(Sanders \& Mirabel 1996), although there is some speculation
that recently discovered faint ``SCUBA'' sources at 
sub-millimetre wavelengths could represent their high-redshift
counterparts (eg. Blain et al. 1999). 
Approximately 80\% of local ULIGS are believed to be powered by starbursts
and the remainder show evidence for an AGN contribution (Genzel et al. 1998; 
Lutz et al. 1998).
Far-infrared observations have shown that dust and molecular
gas in local ULIGS is concentrated in compact regions $\lesssim1$kpc
(Okumura et al. 1991, Bryant 1996) and that a large fraction of the 
optical/UV
emission is hidden by optically-thick dust (Sanders et al. 1988). 
A study of their properties and
importance to  galaxy evolution therefore requires observations
at wavelengths virtually immune to dust absorption. 
Radio frequencies provide an excellent window of opportunity.
 
\begin{figure}
\plotone{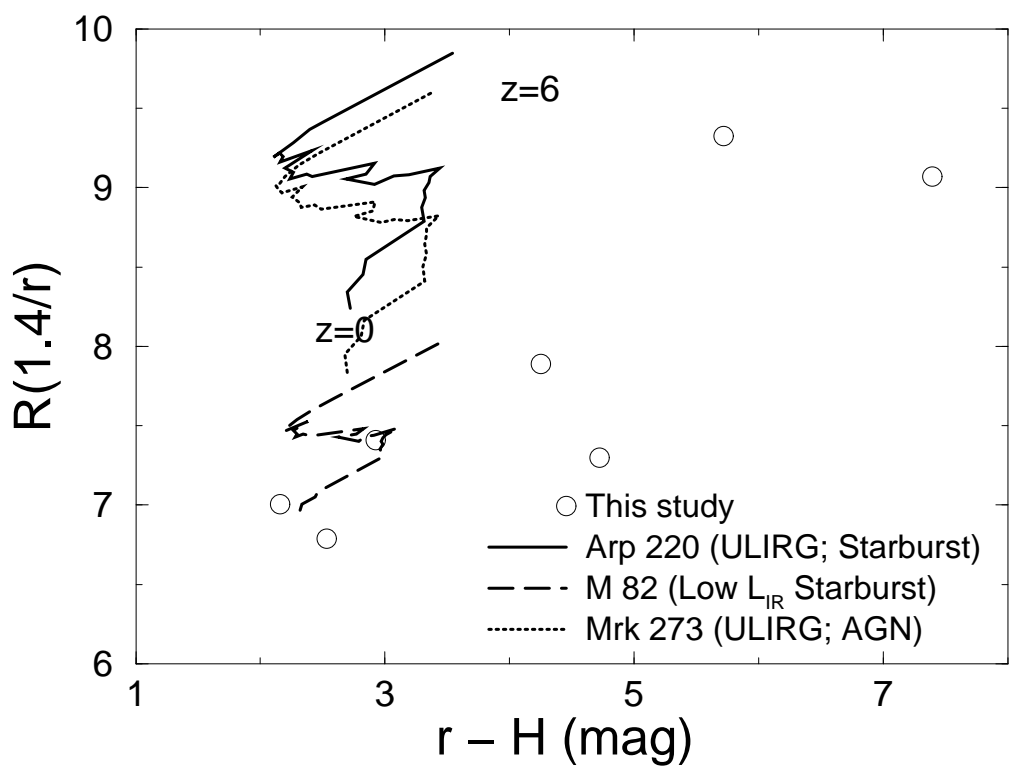}
\vspace{-1.7in}
\caption{Locus of observed radio--to--optical flux ratio versus 
$r-H$ color for $0\leq z\leq 6$ ($\Delta z=0.2$) 
using three local far-IR selected
systems: Arp 220 ($L_{IR}\simeq 1.6\times10^{12}L_{\sun}$; 
strong starburst),
M82 ($L_{IR}\simeq 6\times10^{10}L_{\sun}$; weak/moderate starburst) and 
Mrk 273 ($L_{IR}\simeq2.6\times10^{12}L_{\sun}$; AGN dominated). 
\label{Rkother}}
\end{figure}

Figure~\ref{Rkother} illustrates the predicted locus in color-color space 
using the synthetic SEDs of three local far-IR selected systems:
Arp 220 ($L_{IR}\simeq 1.6\times10^{12}L_{\sun}$) - a ULIG undergoing a 
powerful starburst as seen 
via high resolution radio observations by Smith et al.
(1998); M82 ($L_{IR}\simeq 6\times10^{10}L_{\sun}$) - a system 
undergoing a weak-to-moderate starburst, and Mrk 273 
($L_{IR}\simeq2.6\times10^{12}L_{\sun}$) - a ULIG whose bolometric emission is
believed to be dominated by a hidden central AGN
from the presence of strong Seyfert-2
lines and moderately strong hard X-ray (2-10 keV) emission (Turner et al. 1997).
We have used the synthetic SEDs generated by
Devriendt et al. (1999) to model the starburst emission. 
These authors used a self-consistent modelling approach
to predict the stellar optical/UV/near-IR emission, its reprocessing into 
the mid-IR--to--sub-mm by dust, and the
nonthermal stellar-powered radio emission based on 
the empirical radio--to--far-IR luminosity correlation.
Due to its strong AGN-dominated nature, the starburst synthetic SED predicted
by Devriendt et al. for Mrk 273 differs appreciably from that observed in
the {\it radio}. For this source, we therefore   
used the Devriendt et al. SED at wavelengths $\lambda<1$mm and
extrapolated into the radio using its actual {\it observed} radio--to--1mm spectral
slope and fluxes (obtained from the NASA/IPAC Extragalactic Database
\footnote{The NASA/IPAC Extragalactic Database (NED) is operated
by the Jet Propulsion Laboratory, California Institute of
Technology, under contract with the National Aeronautics and Space
Administration.}).

Figure~\ref{Rkother} shows that a galaxy characteristic of the 
(low-IR luminous) M82 system will occupy a region similar to 
that occupied by normal galaxies in this study (and also their predicted 
synthetic colors in Fig.~\ref{Rr}). Luminous systems classified as 
ULIGS however (Arp 220 and Mrk 273), will tend have higher radio--to--optical
flux ratios which could be easily selected.
This can be explained by the well-observed correlation between 
far-infrared luminosity and 
far-IR ($60-100\mu$m)--to--optical spectral slope
(Soifer et al. 1987). 
Consequently, the most IR-luminous systems with the largest far-IR--to--optical
ratios are also likely to have a high level of radio-emission due to its 
strong correlation with IR luminosity. This will lead to a larger than
average radio--to--optical flux ratio for
ULIGS in general as shown in Fig.~\ref{Rkother}. 
 
The existence of systems with either larger rest-frame optical/UV
extinction or excess AGN contribution to the radio than the ULIGS
considered here will be shifted further upwards on this plot.
Diagnostics to distinguish between AGN and starburst dominated ULIGS
using radio-to-optical color alone will not be trivial and is left to
a future study. The three ULIGS in Fig.~\ref{Rkother} represent a range
of known ULIGS and their location on this plot simply serves as a 
diagnostic to pre-select ULIG candidates for further study.

A system like Arp 220 (with $\nu L_{\nu}(1.4{\rm
GHz})\simeq2.5\times10^{6}L_{\sun}$) 
could be observed to redshift
$z\sim1.6$\footnote{Assumes $q_{0}=0.5$, $H_{0}=50\,{\rm km}\,{\rm
s}^{-1}\,{\rm Mpc}^{-1}$} 
if initially selected from a radio survey
limited to $S_{1.4{\rm GHz}}\simeq50\mu$Jy.  Thus, to limiting
sensitivities reached by existing 1.4 GHz surveys, such a method may
not probe the highest redshifts.  Nonetheless, as shown in
Fig.~\ref{Rkother}, such systems could still be well separated from
normal galaxies to this redshift.  Assuming an Arp 220-like SED and
moderate luminosity evolution ($L_{60\mu{\rm m}}\propto(1+z)^{2.5}$),
the surface density of ULIGS to $z\sim1.6$ is expected to be of order
$150\,{\rm deg}^{-2}$ at $S_{1.4{\rm GHz}}\gtrsim50\mu$Jy, or about
$6\%$ of the integral count to this sensitivity (Richards 2000).  They
should therefore exist in significant numbers in deep large-area
radio surveys.

\section{SUMMARY AND CONCLUSIONS}

We have used the VLA radio telescope to image a contiguous
$33\times33\,{\rm arcmin}^{2}$ area to a (mean) limiting ($5\sigma$) 
sensitivity of $\simeq0.35$ mJy.
From a total of 62 detections, the results of 
optical and near-infrared
photometry are reported for 43 sources. Our optical photometry is more
sensitive than previous optical follow-up studies of radio surveys of
similar depth.
Our main findings are: 

\indent
(1) We have used a robust, likelihood-ratio method for determining optical
identifications and their reliability. 
This method is seldom used in identification
studies and is insensitive to assumptions concerning fluctuations
in background source density and Gaussian error distributions. 
We assigned optical candidates to 26 radio sources with
reliability $\gtrsim80\%$. 
Nine radio sources are uncertain and/or ambiguous, and eight are
empty fields.
Near-infrared photometry from the {\it 2MASS} database was reported 
for 7 sources.

\indent
(2) The eight optical empty field sources all display compact and symmetric
radio morphologies and most probably represent compact starbursts 
at $z\lesssim0.3$ strongly obscured by dust. They may 
require at least 4 magnitudes
optical extinction to account for their 
large radio--to--optical flux ratio compared to the identified population. 
Our conclusion for them being `compact starbursts' is very tentative as it is
purely based on
starburst versus AGN number statistics expected from sub-mJy radio surveys. 
Further deep infrared/optical imaging and spectroscopy will be necessary.

\indent
(3) Consistent with previous studies,
our deep ($r\simeq25$) optical imaging shows that the
optical appearence can be divided into two classes according
to radio flux-density: elliptical-like morphologies for $\gtrsim2$ mJy,
and peculiar or disturbed for $\lesssim2$ mJy.

\indent
(4) Using a stellar synthesis model which includes
radio emission and dust reddening,
we find that the near-infrared--to--optical emission in a small, bright 
sub-sample
is reddened by 
`optically thin'
dust with $A_{V}\simeq2-2.5$ mag, regardless of morphological type.  
This appears consistent with other more direct determinations.
Consistent with previous studies, the radio emission from
early-type systems seems to be powered 
by AGN rather than star-formation to account for their
anomalously large radio--to--optical(--near-infrared) ratios.  

\indent
(5) Our analysis shows that 
a radio/optical or radio/near-IR color selection technique could
provide a potential means for detecting ULIG-type objects
to $z\sim1.6$.  

Despite the lack of spectroscopic information, our  
study of a homogeneous population of faint radio sources has stressed 
the importance of dust on studies of intrinsic galaxy properties
and their evolution at optical 
wavelengths. A future goal would be to obtain spectra, or multi-color
optical/near-infrared photometry to better explore these sources and the 
validity of the simple stellar synthesis models presented in this paper.
The ever improving
resolution (and sensitivity) capabilities of optical/near-IR
detectors over those feasible at (the longest) radio wavelengths requires 
robust identification techniques to better ascertain their properties. 
Likelihood-ratios provide one such technique. 
The present study complements other deep optical studies
of faint radio sources to constitute a statistically significant sample  
for inferring their nature and importance to galaxy evolution.

\acknowledgments

FJM thanks Glenn Morrison and JoAnn O'Linger for valuable assistance
with the data reduction
and Rosalie Ewald for assistance with
radio/optical image overlays. 
We thank the staff at Palomar Observatory for technical
assistance during the observing run. 
This publication makes use of data products 
from the Two Micron All Sky Survey, which is
a joint project of the University of 
Massachusetts and the Infrared Processing and Analysis
Center/California Institute of Technology, 
funded by the National Aeronautics and Space
Administration and the National Science Foundation.
The National Radio
Astronomy Observatory is operated by Associated Universities, Inc.,
under cooperative agreement with the National Science Foundation.
This research has made use of the NASA/IPAC extragalactic database
(NED) which is operated by the jet propulsion laboratory, caltech,
under contract with the national aeronautics and space administration.
FJM acknowledges support from a JPL/NASA postdoctoral fellowship
grant.

\begin{deluxetable}{cccccc}
\tablenum{1}
\tablewidth{0pt}
\tablecaption{The regions used for the source extractions \label{tbl1}}
\tablehead{
\colhead{Region} & \colhead{RA(J2000)\tablenotemark{a}} &
\colhead{Dec(J2000)\tablenotemark{a}} &
\colhead{$\sigma_{rms}$} & \colhead{5$\sigma$ limit} &
\colhead{\# sources\tablenotemark{b}}\nl
\colhead{} &\colhead{} &\colhead{} &
\colhead{(mJy)} & \colhead{(mJy)} & \colhead{}
}
\startdata
1 &$00^h 13^m 48^s.43$ &$+25\arcdeg 48\arcmin 07\arcsec.7$ &0.061 &0.305 &15\nl
2 &$00^h 12^m 34^s.09$ &$+25\arcdeg 47\arcmin 11\arcsec.7$ &0.065 &0.325 &15\nl
3 &$00^h 12^m 33^s.11$ &$+26\arcdeg 03\arcmin 27\arcsec.7$ &0.090 &0.450 &20\nl
4 &$00^h 13^m 48^s.21$ &$+26\arcdeg 02\arcmin 47\arcsec.7$ &0.076 &0.380
&14\tablenotemark{c}\nl
\enddata
\tablenotetext{a}{Defines the center of each region with size
$\approx 16.5\times16.5\,{\rm arcmin}^{2}$.}
\tablenotetext{b}{Total number of sources with flux density
$\geq5\sigma_{rms}$.}
\tablenotetext{c}{Includes the separate components
of two double-component sources.}
\end{deluxetable}

\begin{figure*}
\plotone{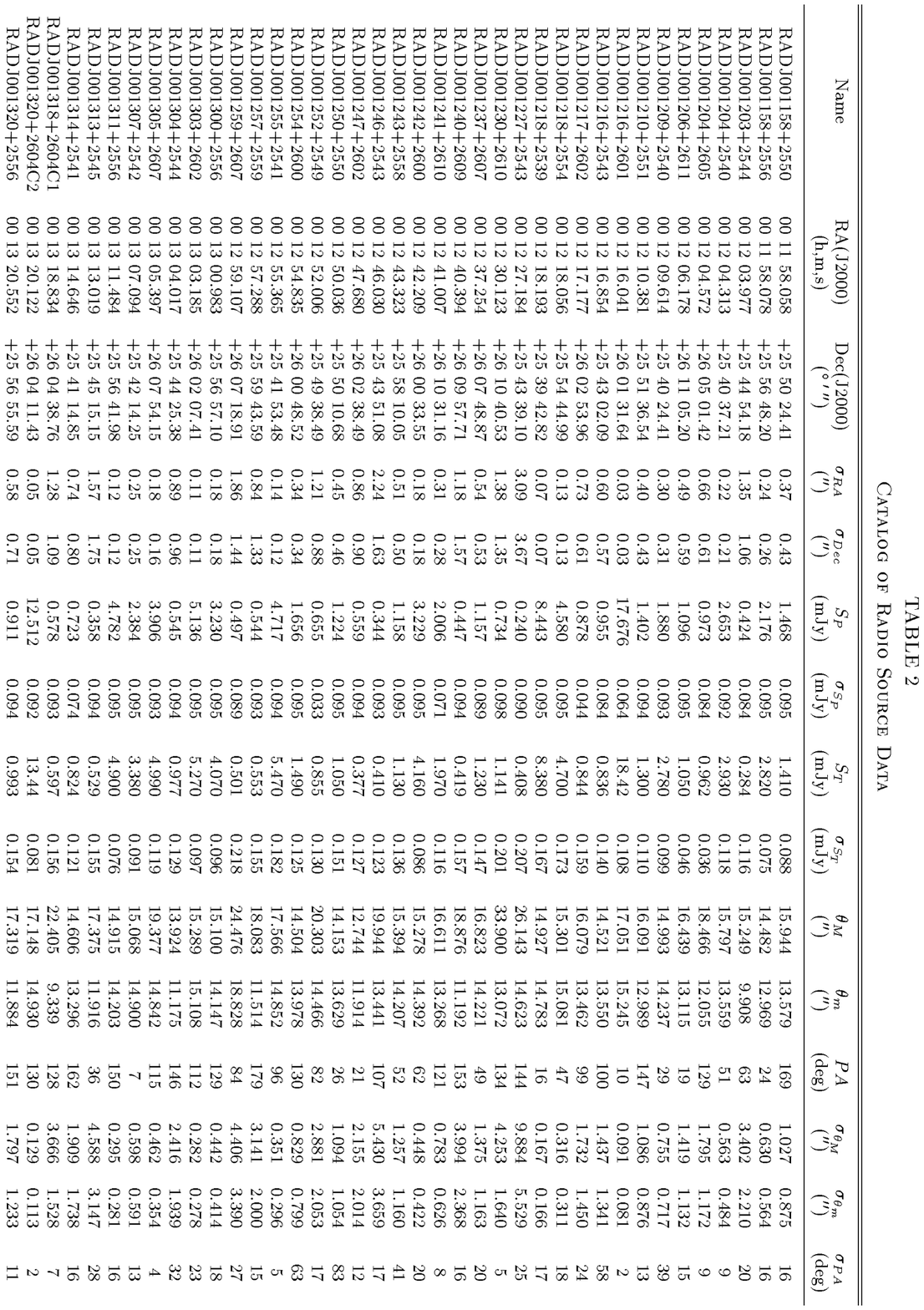}
\end{figure*}

\begin{figure*}
\plotone{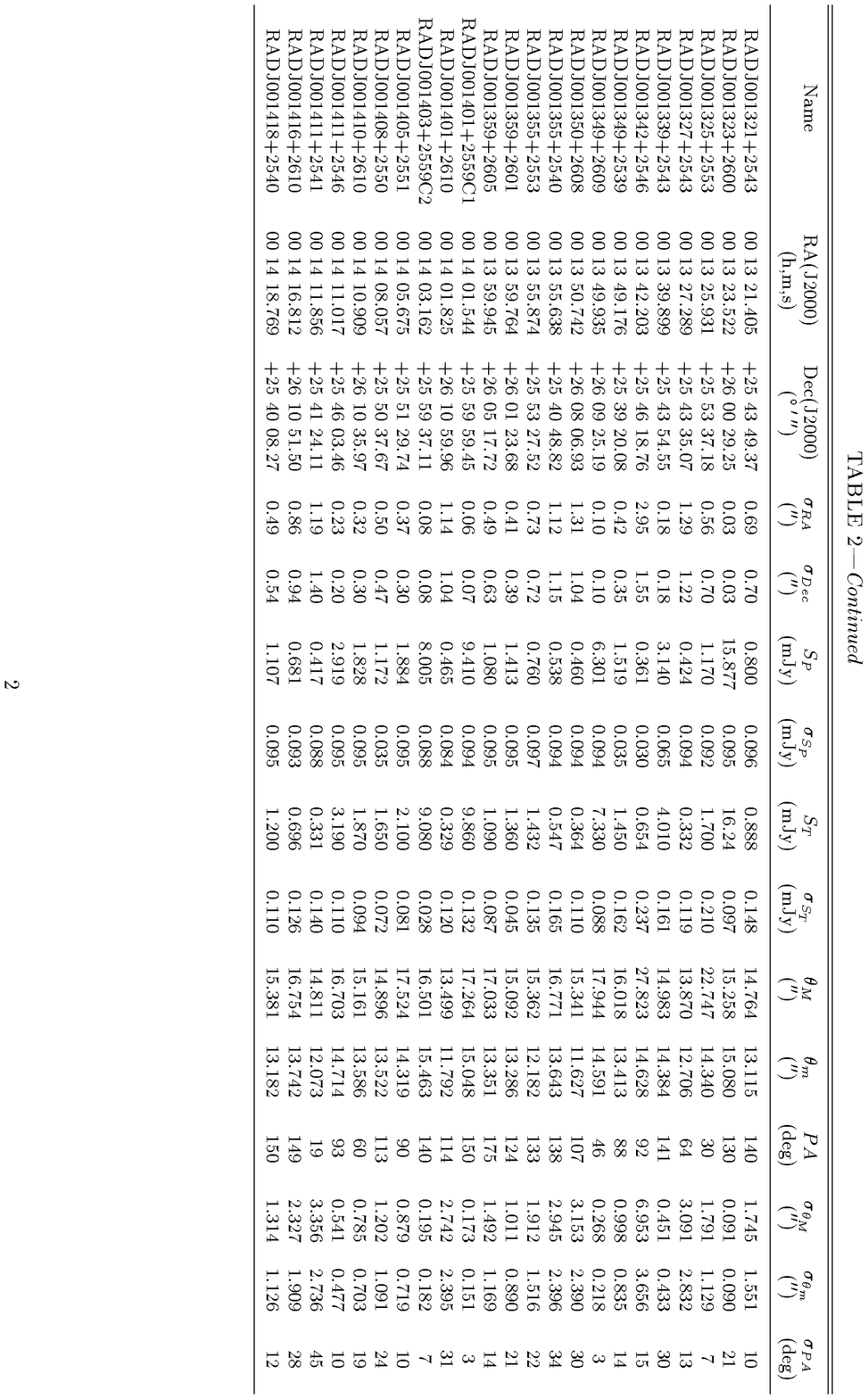}
\end{figure*}

\begin{figure*}
\plotone{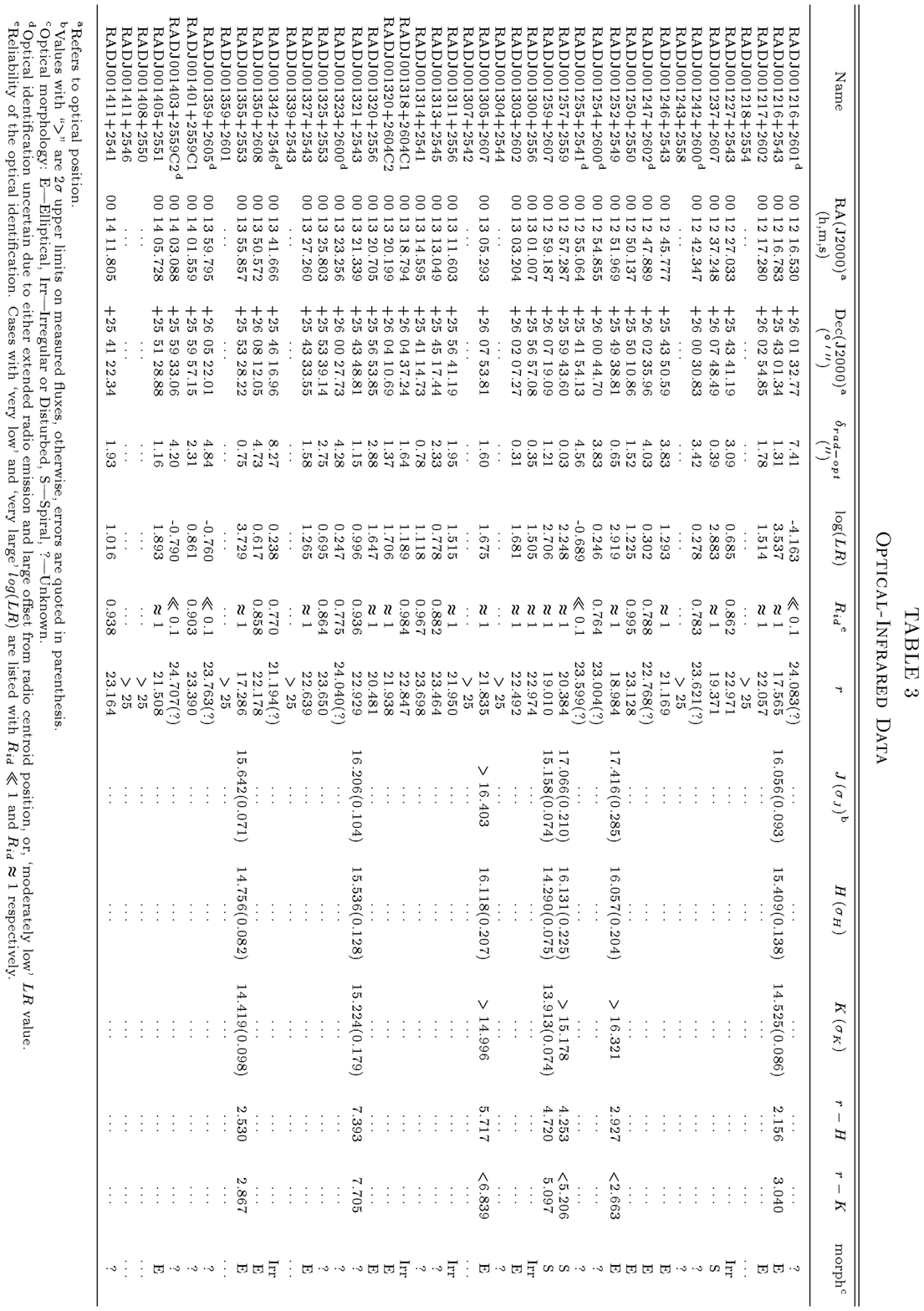}
\end{figure*}

\begin{deluxetable}{cccc}
\tablenum{4}
\tablewidth{0pt}
\tablecaption{Population synthesis models \label{tbl4}}
\tablehead{
\colhead{Type} &
\colhead{SFR($t$)} &
\colhead{IMF} &
\colhead{Age (Gyr)\tablenotemark{a}}
}
\startdata
E/SO &$\tau^{-1}_{1}\exp{(-t/\tau_{1})}$ &Scalo &12.7 \nl
Sab/Sbc &$\tau^{-1}_{8}\exp{(-t/\tau_{8})}$ &Scalo &12.7 \nl
Scd/Sdm &constant &Salpeter &12.7 \nl
SB &constant &Salpeter &0.1\nl
\enddata
\tablenotetext{a}{For the SB-type, an age of 0.1 Gyr
is assumed to apply at all redshifts. For other types, this refers
to the present day age.}
\end{deluxetable}

\end{document}